\documentclass[twocolumn,preprint2,times]{aastex6}

\usepackage{amsmath,amssymb,amsxtra,amsfonts}
\usepackage{graphicx}
\usepackage{color}
\definecolor{midgray}{gray}{0.4}
\definecolor{orange}{rgb}{1,0.5,0}

\shorttitle{DYI: data products and photometric redshifts}
\shortauthors{Liu et al.}
\slugcomment{Draft version}

\begin{document}
\title{Deep CFHT $Y$-band imaging of VVDS-F22 field: I. \\
        data products and photometric redshifts}
\author{Dezi Liu\altaffilmark{1}, 
        Jinyi Yang\altaffilmark{1},
        Shuo Yuan\altaffilmark{1},
        Xue-Bing Wu\altaffilmark{1,2},
        Zuhui Fan\altaffilmark{1},\\
        Huanyuan Shan\altaffilmark{3},
        Haojing Yan\altaffilmark{4},
        and Xianzhong Zheng\altaffilmark{5}
        }
\email{Email: adzliu@pku.edu.cn}
\altaffiltext{1}{Department of Astronomy, School of Physics, Peking University, Beijing 100871, China}
\altaffiltext{2}{Kavli Institute for Astronomy and Astrophysics, Peking University, Beijing 100871, China}
\altaffiltext{3}{Laboratoire d'astrophysique (LASTRO), Ecole Polytechnique F\'ed\'erale 
        de Lausanne (EPFL), Observatoire de Sauverny, CH-1290 Versoix, Switzerland}
\altaffiltext{4}{Department of Physics and Astronomy, University of Missouri, Columbia, MO, USA}
\altaffiltext{5}{Purple Mountain Observatory, Chinese Academy of Sciences, Nanjing 210008, China}

\begin{abstract}
We present our deep $Y$-band imaging data of a two square degree field within the F22 region of 
the VIMOS VLT Deep Survey. The observations were conducted using the WIRCam instrument mounted at
the Canada--France--Hawaii Telescope (CFHT). The total on-sky time was 9 hours, distributed 
uniformly over 18 tiles. The scientific goals of the project are to select faint quasar candidates 
at redshift $z>2.2$, and constrain the photometric redshifts for quasars and galaxies. In this paper, 
we present the observation and the image reduction, as well as the photometric redshifts that we 
derived by combining our $Y$-band data with the CFHTLenS $u^*g'r'i'z'$ optical data and UKIDSS DXS $JHK$ 
near-infrared data. With $J$-band image as reference total $\sim$\,80,000 galaxies are detected 
in the final mosaic down to $Y$-band $5\sigma$ point source limiting depth of 22.86\,mag. 
Compared with the $\sim$\,3500 spectroscopic redshifts, 
our photometric redshifts for galaxies with $z<1.5$ and $i'\lesssim24.0$\,mag have a small systematic offset 
of $|\Delta{z}|\lesssim0.2$, 1$\sigma$ scatter $0.03<\sigma_{\Delta z} < 0.06$, and 
less than 4.0\% of catastrophic failures. We also compare to the CFHTLenS photometric 
redshifts, and find that ours are more reliable at $z\gtrsim0.6$ because of the inclusion 
of the near-infrared bands. In particular, including the $Y$-band data can improve 
the accuracy at $z\sim 1.0-2.0$ because the location of the 4000\AA-break is better
constrained. The $Y$-band images, the multi-band photometry catalog and the photometric 
redshifts are released at \url{http://astro.pku.edu.cn/astro/data/DYI.html}.

\end{abstract}
\keywords{
galaxies: photometry --- 
galaxies: distances and redshifts --- 
astronomical databases: survey --- 
astronomical databases: catalogs
}

\section{Introduction}
Surveys in optical to near-infrared are vital in astronomy.  Due to the characteristics 
of detector response, the conventional charge coupled devices (CCD) used for optical surveys 
are mostly sensitive at $\lambda\lesssim 0.9\,\mu$m (e.g., the Sloan Digital Sky Survey, SDSS; 
\citet{2000AJ....120.1579Y}), 
while the detectors, such as hybrid HgCdTe arrays, for near-infrared 
surveys normally work at $\lambda\gtrsim 1.0\,\mu$m (e.g., 2-Micron All Sky Survey, 
2MASS; \citet{2006AJ....131.1163S}). This leaves a gap of $\sim$\,0.1\,$\mu${m} in between.
The development in detector technology in the past 
decade has allowed this gap to be gradually bridged by extending from both optical and near-infrared. 
Nowadays imaging in this regime is usually done through a broadband filter designated as ``$Y$", 
noting that the shape and amplitude of the spectral response of CCDs in the $Y$-band 
are not identical to their counterparts with near-infrared detectors because of the different structures and fabrications.
Surveys incorporating $Y$-band are now routinely carried out. Some recent examples include the Dark 
Energy Survey (DES; \citet{2014MNRAS.445.1482S}) extending from optical to $Y$ and the UKIDSS 
Large Area Survey (LAS; \citet{2006MNRAS.372.1227D,2007MNRAS.379.1599L}) from $JHK$ to $Y$. 
The uniqueness of $Y$ band has enriched astronomical studies, such as the refinement of quasar 
selection \citep{2010MNRAS.406.1583W} and improving photometric redshift measurements
\citep{2009ApJ...690.1236I, 2014MNRAS.445.1482S}.

Quasars play important roles in studying a variety of subjects ranging from the 
large-scale structure of the Universe to the evolution of galaxies. Being the extreme of active galactic nuclei 
(AGN), quasars are rare objects in the Universe. Thus for quasar studies, an important aspect is to refine 
the candidate selection method such that the final 
sample can be as complete as possible with a low false detection rate at the same time. In this regard, 
the quasar selection in SDSS is highly successful 
(see e.g. \citet{2002AJ....123.2945R}). However, it becomes severely 
incomplete at $z>2.2$. To mitigate this problem, a ``$K$-band excess method" has been 
proposed \citep{2000MNRAS.312..827W,2006MNRAS.367..454H,2008MNRAS.386.1605M}.

In this paper, we present the Deep $Y$-band Imaging project (hereafter DYI; PI: X.-B. Wu), which 
used the CFHT Wide-field InfraRed Camera (WIRCam; \citet{2004SPIE.5492..978P})  to image a two square 
degree area within the VVDS-F22 field. The main scientific aims are to select faint quasar candidates using the 
color criteria proposed by \citet{2010MNRAS.406.1583W} and to study if the photometric redshift 
measurements for quasars and galaxies can be improved.
VVDS-F22 is a contiguous field of VVDS-Wide survey, covering 
4 square degrees \citep{2008A&A...486..683G,2013A&A...559A..14L}. It is also covered 
by SDSS Stripe 82 \citep{2014ApJ...794..120A}, CFHT Legacy Survey 
(CFHTLS), UKIDSS LAS and Deep Extragalactic Survey (DXS), and NRAO Very Large Array Sky 
Survey (NVSS; \citet{1998AJ....115.1693C}). Over 100 quasars at 0.5 $< z <$ 4.7 
within this field have been spectroscopically identified by SDSS and VVDS, and almost all of them (except one) satisfy the 
color-color selection criteria. The DYI photometry, reaching $5\sigma$ point source limiting magnitude of 22.86 mag, 
is 2.0\,mag and 1.5\,mag deeper than 
that of UKIDSS LAS and VISTA Hemisphere Survey (VHS; \citet{2012A&A...548A.119C}), 
respectively, and is comparable to the UKIDSS DXS $K$-band depth.

In addition to quasar selections (Yang et al., in preparation), 
our $Y$-band data can also be important in improving the photometric redshift measurements of the galaxies 
with $z\sim 1.25-1.8$ in the field, because the 4000\AA-break, one of the most prominent spectral 
features that photometric redshift estimate could rely on (e.g. \citet{2009ApJ...690.1236I, 2016ApJS..224...24L}), 
moves to $Y$-band at these redshifts.
Many ongoing and upcoming wide field surveys, especially those for 
weak lensing studies, include $Y$-band observations in order to improve photometric redshift accuracy
beyond $z\sim1.0$, such as DES, 
the Large Synoptic Survey Telescope  (LSST; \citet{2009arXiv0912.0201L}), 
and the Euclid mission \citep{2011arXiv1110.3193L}. As an example, the CFHT Lensing Survey 
(CFHTLenS; \citet{2012MNRAS.427..146H} \& \citet{2013MNRAS.433.2545E}) can only 
obtain robust photometric redshifts ($0.03<\sigma<0.06$) in the range $0.1<z<1.3$ using 
five optical $u^*g'r'i'z'$ bands \citep{2012MNRAS.421.2355H}. 
The LSST, on the other hand, is expected to reach $\sigma \sim 0.02$ over $0<z<3$, 
and one of the main reasons that enable such an accuracy at $z<2$ is the inclusion of $Y$-band.

Reduction of ground-based $Y$-band data, however, is non-trivial 
\citep{1992MNRAS.259..751R, 2010PASP..122..722H}. 
In this paper, we present the DYI observation and image reduction.
We also show the photometric redshift measurements for galaxies in 
combination with CFHTLenS and UKIDSS DXS imaging data. 
In particular, we analyze the impact of $Y$-band photometry on photometric redshift measurements.

The paper is organized as follows. In Section \ref{sec:obs}, 
we introduce the observations of our DYI project. The detailed data reduction procedures are 
presented in Section \ref{sec:red}. In Section \ref{sec:mphot}, we describe the multi-band photometry 
and the treatment of the PSF inhomogeneity between different band images. 
We present the derivation of photometric redshifts  for galaxies and the analysis of their accuracy in Section \ref{sec:photoz}. 
Summaries are presented in Section \ref{sec:sum}. Note that all magnitudes in this paper 
are in the AB system. The conversion constants between the Vega and AB magnitudes for UKIDSS 
$YJHK$ bands are from \citet{2006MNRAS.367..454H}.

\section{Observation}\label{sec:obs}
\begin{figure*}
\centering
\includegraphics[width=0.9\textwidth]{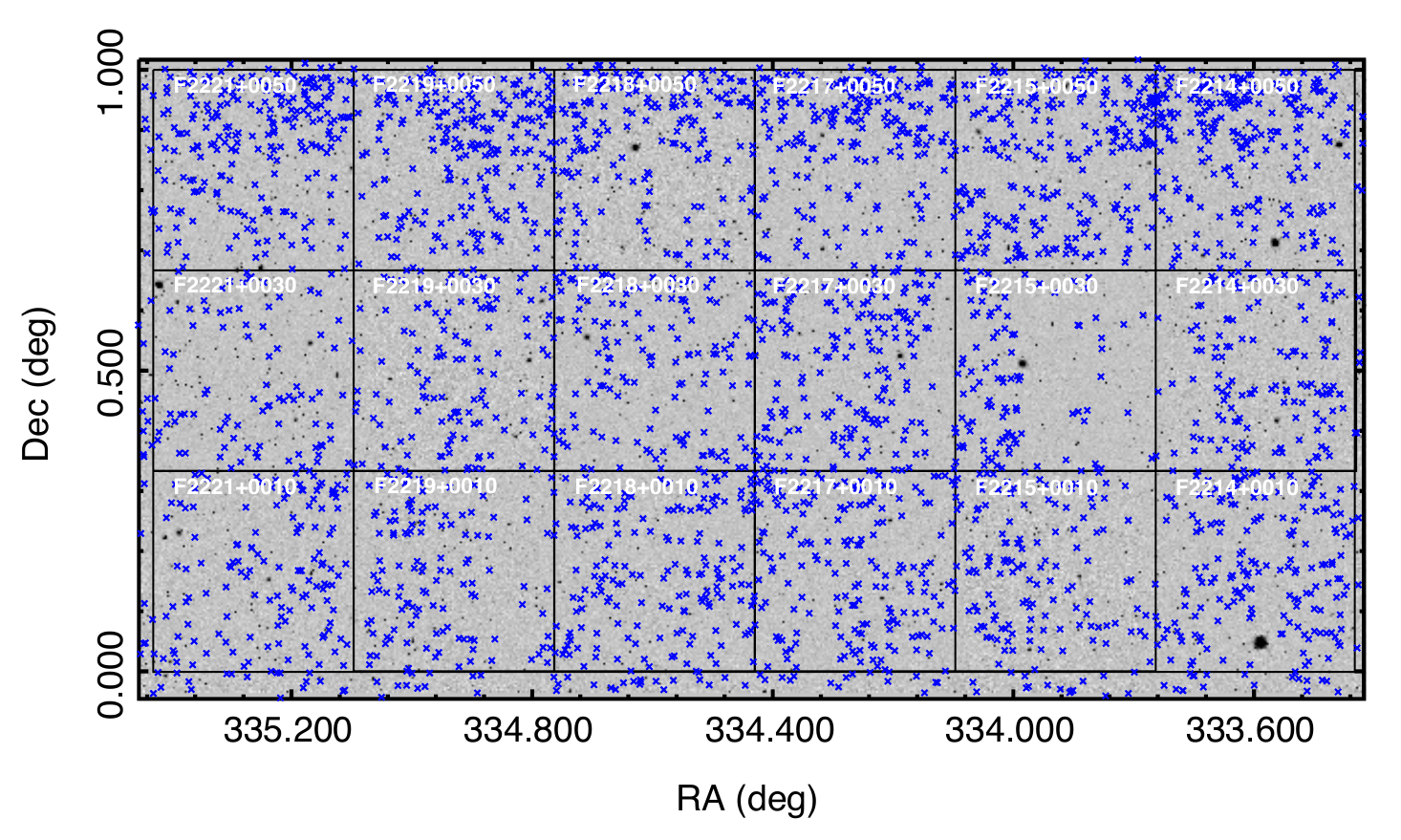}
\caption{Sky coverage ($128\arcmin \times 65\arcmin$) of the final, stacked $Y$-band image.
The 18 tiles, each $20\arcmin \times 20\arcmin$ in size, 
are overlaid as the thin, black grids. Blue crosses indicate the positions of galaxies with spectroscopic redshifts.
The background image is shown with inverted grey scale, and the large black points represent the saturated stars.}
\label{fig:dyian}
\end{figure*}

The DYI project was conducted during August to October 2012
using the WIRCam instrument mounted on CFHT. 
The WIRCam focal plane is made of a mosaic of four HAWAII2-RG detectors, each containing 
$2048\times2048$ pixels, with a sampling of $0.3\arcsec$ per pixel. 
Each detector is read in parallel using 32 amplifiers.
The detectors are operated at $\sim$\,80 Kelvin 
with the readout noise of $\sim$\,30\,e$^-$  and very low dark current ($\sim$\,0.05\,e$^-$/sec).
The average electronic gain is 3.8\,e$^-$/ADU. 
The field of view of the full mosaic is about $21.5\arcmin \times 21.5\arcmin$. 
The whole DYI field was divided into 18 uniform tiles, each $20\arcmin\times20\arcmin$ in size. 
Figure \ref{fig:dyian} shows this layout on top of the final mosaic, which amounts to 
$128\arcmin \times 65\arcmin$. Blue crosses 
indicate the positions of galaxies with reliable spectroscopic redshifts.
Our observation, carried out in Queue Service Observing mode (QSO), took place over two periods: 
one was in August with single exposure time of 127.5 
seconds, and the other spanned the period from September to October with 
single exposure of 115.0 seconds. The total exposure time of each tile
is about 0.5 hour.

\begin{figure}
\includegraphics[width=0.45\textwidth]{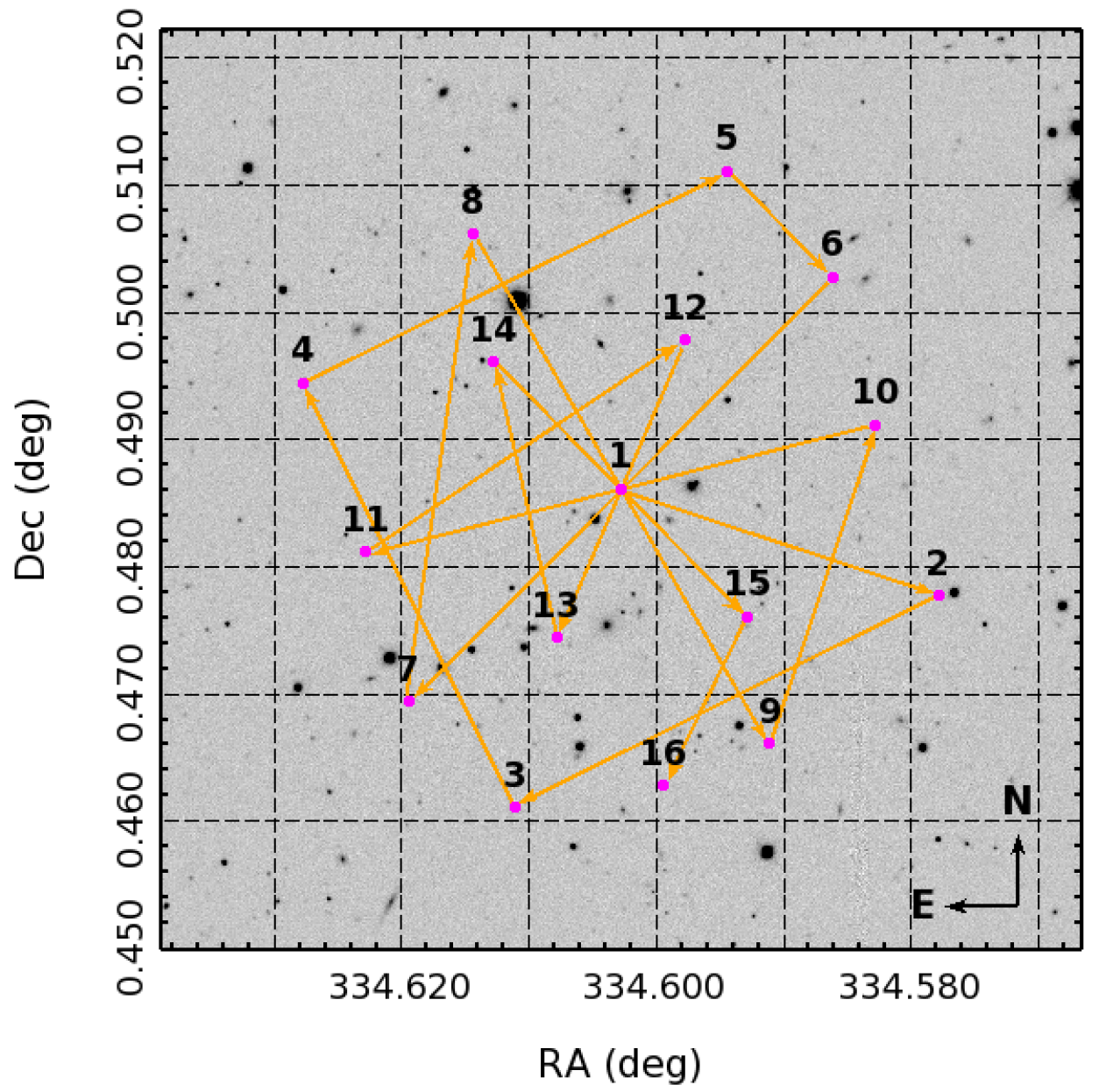}
\caption{The dither pattern of the tile F2218+0030. The mosaic shows the inner
$4.3\arcmin\times4.3\arcmin$ area centered at the first exposure. The magenta 
dots represent the central positions of the 16 dithered exposures, while the arrows
indicate the dither directions and amplitudes between two adjacent exposures.}
\label{fig:dither}
\end{figure}

For each tile we requested 16 dithered exposures in order to fill the $45\arcsec$ inner gaps
between the detectors. In addition, the dither mode can not only improve the spatial sampling, 
but also avoid the detector defects (e.g., the bad pixels) and efficiently remove cosmic rays and the 
high sky background. Figure \ref{fig:dither} shows the dither pattern for the tile F2218+0030 with 
magenta dots denoting the central celestial coordinates of the 16 exposures. The arrows in the 
figure indicate the dither directions and amplitudes between two adjacent exposures. On average 
the dither shift of each tile is about $1.8\arcmin$.

\section{Image Reduction}\label{sec:red}
The WIRCam images are graded at five levels based on the seeing condition, 
sky background, timing etc., and we only use Grade 1 and 2 images, 
which are good for scientific purpose. The raw images were 
preprocessed by 'I'iwi 
pipeline\footnote{\url{http://www.cfht.hawaii.edu/Instruments/Imaging/WIRCam/IiwiVersion2Doc.html}}
(version 2.0), including flagging the saturated pixels, non-linearity correction, reference pixels subtraction, 
dark subtraction, flat fielding, bad pixels and guide window masking, and basic cross-talk removal. 
In the following work, we will concentrate on these preprocessed images, 
referred as \textit{science images}. We note that 'I'iwi pipeline also provided the sky subtracted
images, but in this work we perform the sky subtraction using our own method.
Table \ref{tab:dyi} summarizes the basic observational information. The exposure time of 
each tile used in the work is listed in the seventh column. These exposures are selected to
be continuous to ensure that the seeing does not vary too much. The seeing is the median value of all useful
exposures, ranging from 0.56$\arcsec$ to 1.18$\arcsec$. We note that almost one third of the images have seeing
larger than 1.0\arcsec. For the sake of sensitivity, we do not further reject the science images on the basis of their 
seeing values. Three tiles, F2215+0050, F2217+0050 and F2218+0050, were observed under thin cirrus,
but this did not affect their photometric zeropoint determination when following the same 
reduction procedures as other photometric tiles. Therefore, they are included in our process.

\subsection{Cosmic Rays and Satellite Track Removal}
Even with 16 exposures, the rejection of defects such as cosmic rays could still be 
non-optimal if we rely on the conventional algorithm such as sigma-clipping. To get the best 
co-adds, we opt to remove such defects from the single images as much as possible.

The cosmic rays are detected and removed. This is done by a \texttt{Python} 
code\footnote{\url{http://obswww.unige.ch/~tewes/cosmics_dot_py/}} that implements 
the L.A. Cosmic algorithm, which is based on a variant of Laplacian edge detection \citep{2001PASP..113.1420V}.
The method is capable of rejecting cosmic rays of arbitrary shape by applying a 2D Laplacian convolution 
kernel which is sensitive to variations on small scales. In addition, saturated stars and their full saturation 
trails can also be flagged by the module automatically. In some cases, however, it misclassifies the 
peaks of unsaturated bright stars as cosmic rays. We modify the module to suite the work and find the 
best parameters by visual inspection of the mask images. The rejection is done through three iterations, 
and the results are satisfactory.

In general, the sky background dominates the near-infrared images so that the satellite trails are relatively
too faint to be detected. Instead, we identify the satellite trails from the 
sky subtracted images produced by 'I'iwi pipeline and then mask them from the science images.

\begin{deluxetable*}{cccrrrrr}
\tablewidth{0pt}
\tablecaption{Summary of Our DYI Project. \label{tab:dyi}}
\tablehead{\colhead{Field ID} & \colhead{R.A.}    & \colhead{Dec.} & \colhead{Month, Year} & 
           \colhead{N$_\mathrm{tot}$} & \colhead{N$_\mathrm{eff}$} & \colhead{Exp$_\mathrm{eff}$} & \colhead{Seeing}\\
 & \colhead{(J2000)} & \colhead{(J2000)} &  &  & & \colhead{(s)} &  \colhead{($\arcsec$)}}
\startdata
F2214+0010 &  22:14:28 & +00:10:00 & Aug.-Sep.,2012 & 16 & 16 & 1840.0 & 0.74 \\
F2214+0030 &  22:14:28 & +00:30:00 & Oct.,2012      & 16 & 16 & 1840.0 & 1.07 \\
F2214+0050 &  22:14:28 & +00:50:00 & Oct.,2012      & 18 & 16 & 1840.0 & 1.09 \\
F2215+0010 &  22:15:48 & +00:10:00 & Aug.,2012      & 19 & 7 & 892.5 & 0.57 \\
F2215+0030 &  22:15:48 & +00:30:00 & Aug.,2012      & 16 & 16 & 2040.0 & 0.79 \\
F2215+0050 &  22:15:48 & +00:50:00 & Sep.,2012      & 16 & 11 & 1402.5 & 0.70 \\
F2217+0010 &  22:17:08 & +00:10:00 & Aug.,2012      & 16 & 16 & 2040.0 & 0.56 \\
F2217+0030 &  22:17:08 & +00:30:00 & Aug.,2012      & 16 & 16 & 2040.0 & 0.79 \\
F2217+0050 &  22:17:08 & +00:50:00 & Sep.,2012      & 16 & 16 & 2040.0 & 0.80 \\
F2218+0010 &  22:18:28 & +00:10:00 & Aug.,2012      & 16 & 16 & 2040.0 & 0.57 \\
F2218+0030 &  22:18:28 & +00:30:00 & Aug.,2012      & 16 & 16 & 2040.0 & 0.75 \\
F2218+0050 &  22:18:28 & +00:50:00 & Sep.,2012      & 10 & 7 & 892.5 & 0.75 \\
F2219+0010 &  22:19:48 & +00:10:00 & Oct.,2012      & 15 & 13 & 1495.0 & 1.15 \\
F2219+0030 &  22:19:48 & +00:30:00 & Oct.,2012      & 16 & 14 & 1610.0 & 0.78 \\
F2219+0050 &  22:19:48 & +00:50:00 & Oct.,2012      & 16 & 15 & 1725.0 & 0.74 \\
F2221+0010 &  22:21:08 & +00:10:00 & Oct.,2012      & 16 & 16 & 1840.0 & 0.67 \\
F2221+0030 &  22:21:08 & +00:30:00 & Oct.,2012      & 16 & 16 & 1840.0 & 1.18 \\
F2221+0050 &  22:21:08 & +00:50:00 & Oct.,2012      & 16 & 16 & 1840.0 & 1.15
\enddata
\tablecomments{The fifth column shows the number of raw exposures of each tile, while the sixth column 
N$_\mathrm{eff}$ represents the number of images with \texttt{GRADE} 1 and 2. The total 
exposure time of each tile used in the work is listed under Exp$_\mathrm{eff}$. The quoted seeing value
in the last column is the median of all good exposures. Three fields, 
2215+0050, 2217+0050 and 2218+0050, were observed under non-photometric condition.}
\end{deluxetable*}

\subsection{Sky Background Subtraction}
Due to the OH emission lines, the $Y$-band sky background varies rapidly, 
at a time scale of a few minutes to an hour \citep{1992MNRAS.259..751R, 2010PASP..122..722H}. 
As a compromise between the limited exposure times and the temporal variations of the sky, 
we perform the background subtraction on images with continuous exposure time no more than 15 
minutes for each tile. Therefore, if the exposure is longer than that, the tile will be divided into
two roughly uniform groups.

For each group, we run \texttt{SExtractor} (version 2.19.5; \citet{1996A&AS..117..393B}) 
to create the preliminary background 
subtracted images. Then they are integer-registered to a common grid by matching dozens 
of stars with high signal-to-noise ratios, and stacked to an intermediate image. 
All objects are extracted from the stacked image by applying a low detection threshold and masked 
from the original science images. In general, the background values of the mask regions
can be estimated by two dimensional interpolation with nearby pixels. However, it leads to some obvious
artificial features in the large mask regions, especially at the positions of bright objects. 
Moreover, performing two dimensional interpolation is relatively time-consuming.
For a small blank region in the image, we find that the distribution of the pixel values can be well fitted by
a Gaussian function. Therefore, for a specific mask region, the median value $\bf{a}$ and variance
$\bf{\sigma^{2}}$ are calculated through its adjacent pixels (at least 900 unmasked pixels)
to construct a Gaussian pseudo-random number generator following distribution $N(\bf{a}, \bf{\sigma^2})$.
The masked pixels are filled with random numbers sampled by the generator. This procedure
conserves the local statistical properties, and avoids many artificial effects due to the interpolation. 
Finally, the background map of a certain exposure is created by a running median of all other masked 
images. As an example, for a group having 8 exposures, the background map of the third exposure is created 
by combing other seven exposures. During the procedure, we apply $3\sigma$-clipping to reject the bad pixels.

\subsection{Other systematics Removal}
After the background subtraction, the images still show visible systematics that appear as horizontal 
stripes because of the residual amplifier differences. Similar to \citet{2014ApJS..210....4M}, 
we take the median along the $x$-axis for the pixels belonging to a given amplifier. The variation 
is smoothed and fitted with a linear equation. We estimate the best fitted coefficients through 
least squares method and subtract this trend from the images.

We do not see any obvious large scale variations in the images as mentioned by 
\citet{2014ApJS..210....4M}. However, the images of the forth detector indeed show very 
small inhomogeneities, and the patterns are not constant between different exposures. 
The amplitude is indistinguishable from the variance of the background noise so 
that it is not expected to affect the photometry. Therefore, we do not further take the 
inhomogeneities into account.

\subsection{Astrometric and Photometric Calibrations}
We use \texttt{SCAMP} (version 2.2.6, \citet{2006ASPC..351..112B}) for 
astrometric calibration with 2MASS point source catalog as reference.
To find the accurate astrometric solution, the \texttt{SCAMP} is run twice. 
We consider the detector positions to be independent between exposures 
(\texttt{MOSAIC\_TYPE = LOOSE}) for the first time. In that case, the astrometric 
calibration is conducted separately for each exposure. Then we run \texttt{SCAMP} 
again to derive a common and median relative positioning of the four detectors within 
the focal plane (\texttt{MOSAIC\_TYPE = FIX\_FOCALPLANE}). The final derived 
pixel scale varies radially in a symmetric way by 0.6\% from the center to the 
edge of the field of view. The rms offsets of the astrometry, comparing with the 2MASS 
world coordinates, are less  than 0.14$\arcsec$ along both right ascension and declination axes.

Because the DYI field has been covered by UKIDSS LAS $Y$ band 
(hereafter $Y_{\mathrm{LAS}}$),
we use its photometry as reference to perform photometric calibration. 
The SDSS Stripe 82 standard star catalog \citep{2007AJ....134..973I} is 
cross-matched with  $Y_{\mathrm{LAS}}$-band catalog to get the $Y_{\mathrm{LAS}}$-band 
magnitudes. We further constrain the stars to have signal-to-noise ratios larger than 10 in 
$Y_{\mathrm{LAS}}$ band (corresponding magnitudes brighter than 19.0\,mag). In total, 
there are about 2500 stars that uniformly distribute in the field.

The effective wavelength and bandpass of  $Y_{\mathrm{LAS}}$ band are 1.03\,$\mu$m 
and 0.1\,$\mu$m, respectively. Though it is quite similar to WIRCam $Y$ band, we also 
derive a color correction between the two bands by including UKIDSS LAS $J$ band in 
order to avoid any potential magnitude bias.
The Pickles star spectra library \citep{1998PASP..110..863P} is used to 
calculate the convolutional magnitude. The linear color relation can be written as
\begin{eqnarray}
\label{eq:color}
Y - Y_{\text{LAS}} = 0.0342 \times (Y - J)_{\text{LAS}} + 0.0042
\end{eqnarray}
with rms scatter of 0.004\,mag. As expected, the coefficients in the color relation 
are very small and the color correction is at most 0.03\,mag.
Then we determine the zeropoint by the equation
\begin{eqnarray}
m_\text{zp} = m_{Y} + 2.5 \times \log{f} + c - \alpha \times {k_{\nu}},
\end{eqnarray}
where $m_{Y}$ represents the apparent $Y$-band magnitude of each star
and $f$ is the corresponding observed flux in unit of $\mathrm{ADU\,s^{-1}}$.
The parameter $c$ is the color correction term in equation (\ref{eq:color}) and
 $\alpha$ and $k_{\nu}$ are the airmass and corresponding coefficient. 
To determine $k_{\nu}$, we extrapolate the Extinction Curve of Mauna Kea 
from the recent results by \citet{2013A&A...549A...8B} to 1.0 $\mu$m.

However, we note that the zeropoints are different between the four WIRCam detectors due to the different gain 
values, as also pointed out by \citet{2014ApJS..210....4M} when analyzing $K_s$-band images. 
The offsets can be as large as 0.1\,mag. Therefore, we do photometric calibration for 
each detector separately, and then correct the offsets to a common zeropoint. Specifically, for a certain 
tile we run firstly \texttt{SCAMP} on the images of each detector to homogenize the flux scales 
(i.e. parameter \texttt{FLXSCALE} in headers). \texttt{SWarp} (version 2.38.0; 
\citet{2002ASPC..281..228B}) is then used to median stack the images with 
resampled pixel scale of $0.3\arcsec$. We extract the source catalog by \texttt{SExtractor} and 
cross match with the reference star catalog. The median value of the differences 
between the instrumental and apparent magnitudes is calculated as the detector's zeropoint.
Finally, we set the zeropoints of the four detectors to 24.30\,mag by rescaling 
the flux scale parameter in the image headers.

\begin{figure}
\includegraphics[width=0.5\textwidth]{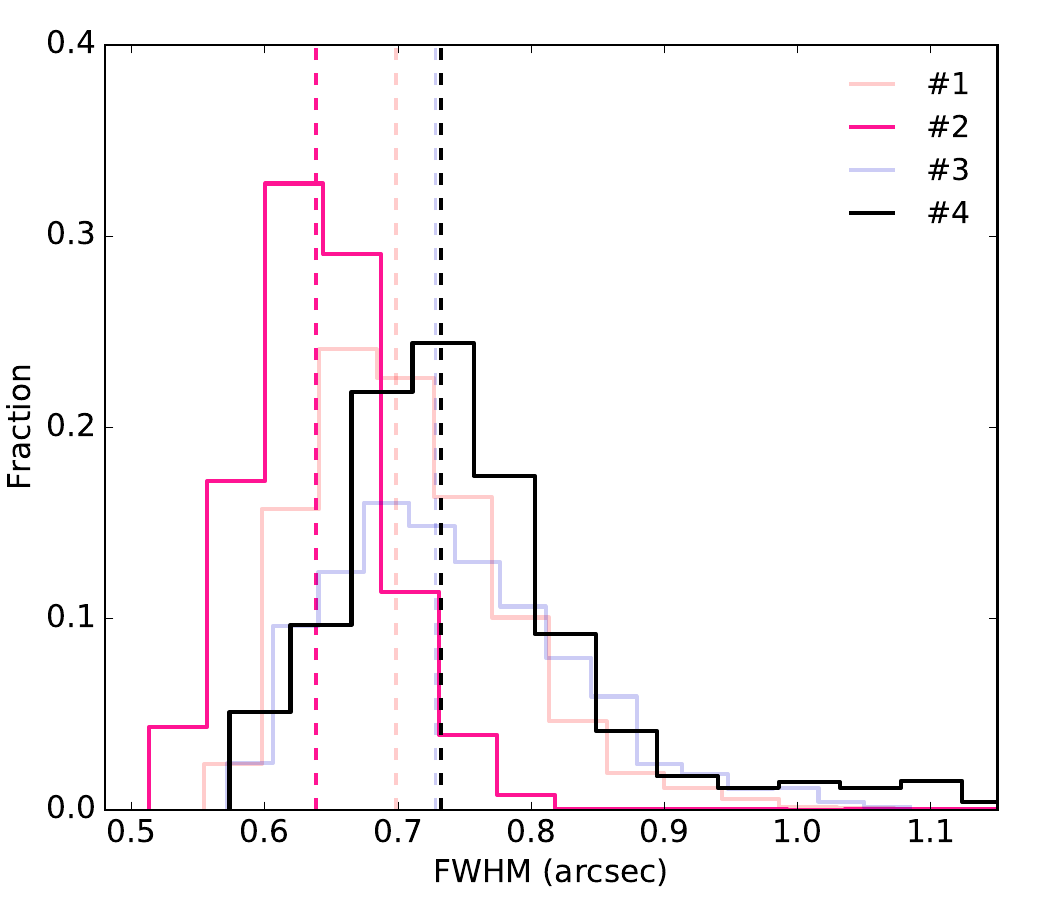}
\caption{Fractional FWHM histograms of the four detectors for tile F2217+0010. 
Vertical dashed lines indicate the median FWHM values. To emphasize the 
difference between detector 2 and 4, the color contrasts of the other two detectors are downweighted.
The FWHM is estimated by \texttt{SExtractor} for stars with signal-to-noise ratios larger than 20.}
\label{fig:see}
\end{figure}

\subsection{Image Stacking and Photometric Quality}
We stack the background subtracted images using \texttt{SWarp} in combination with
the astrometric and photometric calibrated headers calculated by \texttt{SCAMP}. 
However, after stacking all the images into one mosaic for a certain tile, the distribution 
of the full width at half maximum (FWHM) of stars shows either multiple peaks or large dispersion 
of over $0.3\arcsec$. This problem will definitely bias the aperture photometry. 
By checking the FWHM distribution of the mosaic for each detector, however we find that 
it is well behaved which shows small variation, typically within $0.2\arcsec$. In other words, 
the FWHM distributions are slightly different between the detectors. As a result, it leads to the 
FWHM distribution of the stacked image having multiple peaks or large dispersion. 
The difference presumably results from the different responses of the four detectors, 
the dither mode, or the image stacking procedures. However, for the last two cases, it is not 
supposed to see any deviation from the FWHM distributions of individual exposures. 
Therefore, we count the FWHM values of stars in every single exposure, and then analyze their
statistical distributions, by combining these values of all exposures, of the four detectors individually. 
Figure \ref{fig:see} shows an example for tile F2217+0010. The FWHM is estimated by 
\texttt{SExtractor} for stars with signal-to-noise ratios larger than 20.
From the figure we can see clear deviations even in the individual exposures, 
of which the median offset between detector 2 and 4 is nearly $0.1\arcsec$. Further, 
we calculate the median FWHM value of every exposure for a given detector in this tile.
We find that the FWHM values for detector 2 are almost homogeneous to better than $0.1\arcsec$,
compared to detector 4. Based on above analysis, the difference between the detectors is 
most probably attributed to the different detector features. However, the details need to be 
further studied and beyond the scope of this paper.

On the other hand, the FWHM values between different exposures for each detector are similar.
Consequently, we stack the dithered images of each detector individually using median 
combination method, and then perform point spread function (PSF) homogenization in the subsequent 
section to obtain the final mosaic. To stack the images of each detector, we split the \texttt{SCAMP} header of every 
exposure into four sub-headers which contain the calibrated astrometry for individual detectors 
\footnote{The header of each exposure generated by \texttt{SCAMP} actually contains the astrometric 
solution for each detector independently. Therefore, we can extract the astrometric solution for each 
detector from the header without losing any information.}.
Then the images for each detector can be stacked using \texttt{SWarp} in combination with the corresponding 
sub-headers. Such realization ensures that the astrometric calibration is homogeneous between the 
detectors. The dithering strategy allows us to further remove some remanent instrumental defects and 
cosmic rays during the stacking procedures.
The pixel scale is resampled to $0.186\arcsec$, using Lanczos3 interpolation function, to match with 
CFHTLenS optical images. The stacked image covers a sky area of about 
$11.26\arcmin\,\times\,11.26\arcmin$. Because any two adjacent 
images are partially overlapped due to the dithered exposures, the total coverage of each tile is about 
$22.25\arcmin\,\times\,22.25\arcmin$.

\begin{figure}
\includegraphics[width=0.5\textwidth]{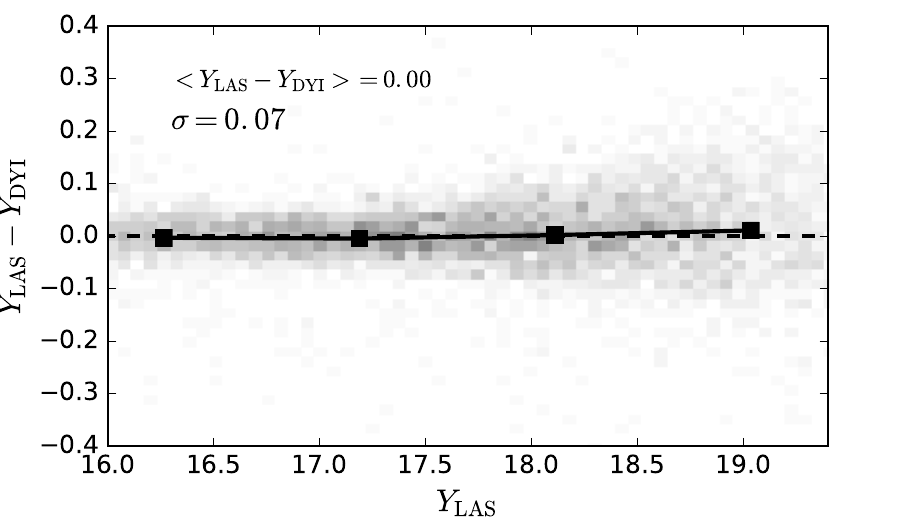}
\caption{Photometric calibration accuracy of $Y$-band data versus UKIDSS LAS data. 
The grey points represent stars cross-matched with SDSS standard star catalog in Stripe 
82 field with signal-to-noise ratio larger than 10 in both $Y$ bands. 
The mean differences in given magnitude bins, shown as black squares, are stable and 
close to zero.}
\label{fig:magdiff}
\end{figure}

We measure the magnitudes of the Stripe 82 standard stars
from the stacked images using \texttt{SExtractor} to check the photometric quality. 
Figure \ref{fig:magdiff} shows the comparison between the measured total magnitudes 
(\texttt{MAG\_AUTO}) and $Y_{LAS}$-band magnitudes. The stars are selected to have 
signal-to-noise ratios larger than 10 in both $Y$-band catalogs. The dispersion is nearly Gaussian
with $\sigma \sim 0.07$\,mag,
confirming that our background subtraction and calibration procedures are valid.
The mean differences are calculated in four magnitude bins, shown as black squares in the figure. 
It can be seen that the trend as a function of magnitudes is stable and close to zero.

\section{Multi-band Photometry}\label{sec:mphot}
At mentioned above, the VVDS-F22 field has also been covered by various surveys, such as
SDSS,  CFHTLS, UKIDSS LAS \& DXS, and NVSS, from optical to radio.
To construct our multi-band photometry catalog in the DYI field, we opt to incorporate 
the CFHTLenS data in optical \citep{2012MNRAS.427..146H, 2013MNRAS.433.2545E}
and the UKIDS DXS data in near-infrared \citep{2007MNRAS.379.1599L} in this field, respectively, 
as they are the deepest. Our catalog is based on our own photometry in these nine bands: 
$u^*$, $g'$, $r'$, $i'$, $z'$ (CFHTLenS); $Y$ (DYI); and $J$, $H$, $K$ (DXS). 
Table 2 summarizes the basic characteristics of these data. We note that the $H$-band data only 
cover $\sim$\,39\% of the DYI field. We discuss our photometry in detail below.

\begin{deluxetable}{ccccl}
\tablewidth{0pt}
\tablecaption{Summary of Multi-Band Data. \label{tab:f22}}
\tablehead{\colhead{Band} & \colhead{FWHM[$\arcsec$]} & 
\colhead{Zeropoint} & \colhead{5$\sigma$ Depth\tablenotemark{a}} & \colhead{Survey}}
\startdata
$u^*$ & 0.72-1.03  & 25.16 & 25.08 & CFHTLenS \\
$g'$  & 0.67-0.84  & 26.37 & 25.49 & CFHTLenS \\
$r'$  & 0.61-0.73  & 25.96 & 25.33 & CFHTLenS \\
$i'$  & 0.53-0.71  & 25.65 & 24.59 & CFHTLenS \\
$z'$  & 0.48-0.79  & 24.73 & 23.28 & CFHTLenS \\
$Y$   & 0.56-1.33  & 24.30 & 22.86 & CFHT DYI  \\
$J$   &  0.69-1.10 & 25.45 & 23.51 & UKIDSS DXS \\
$H$   & 0.64-1.08  & 26.10 & 23.15 & UKIDSS DXS \\
$K$   & 0.55-1.32 & 25.94 & 23.20 & UKIDSS DXS
\enddata
\tablenotetext{a}{The $5\sigma$ depth is measured within a circular aperture of 2.2$\arcsec$
in diameter.}
\end{deluxetable}

\subsection{Image Alignment and PSF Homogenization}
When combining multi-band data for photometric redshift measurements, it is critical to carry out 
photometry homogeneously across all bands to obtain accurate color information. 
The most common treatment is to align all the images to the same grid and then perform 
the so-called ``matched-aperture photometry¡±, i.e.,  using the same aperture for a given 
object in all bands. We adopt this approach, and prepare our images as follows.

First, all the CFHTLenS and the DXS images are aligned to our $Y$-band images by using 
\texttt{SWarp}. In this process, the images are resampled to the pixel scale of 0.186\arcsec
using Lanczos3 interpolation function, and projected to the tangential 
planes with maximum positional error of 0.001 pixel.

The next step is to homogenize the different PSF sizes among these images, 
which is necessary because the point-source FWHM values vary significantly 
among them. If this is not treated, using the same aperture for a given source in 
all bands would mean loosing different fraction of total light across the bands, which 
in turn would result in large systematic errors in the colors and consequently in  
photometric redshift measurements.

The most straightforward method of PSF homogenization is to degrade all the images 
to the PSF size of the image that has the worst point-source FWHM 
(e.g., \citet{2009ApJS..183..295T,2010ApJS..189..270C}).  The assumption is that the 
PSF of a given image is spatially invariant, which is a reasonable approximation in our 
case. For the sake of simplicity, we chose to degrade all images to the PSF size of the 
$Y$-band image that has the worst quality, which is the forth detector in tile F221+0330 
with FHWM of 1.33\arcsec.

We constructed an empirical PSF for each image. To select the candidate PSF stars, 
we first run \texttt{SExtractor} to generate a catalog of bright sources, and retain
only point-like sources based on the magnitude versus size diagram. This catalog is
then matched to the SDSS Stripe 82 standard stars for refinement.
To avoid any possible nonlinearity, we further restrict the peak pixel values
of the selected stars to be less than half of the saturation level. 
The peak positions are then shifted to a common center by 
applying two dimensional cubic interpolation.
We normalize the stars to their total fluxes, and
perform median stack to construct the PSF for each image.

\begin{figure}
\includegraphics[width=0.5\textwidth]{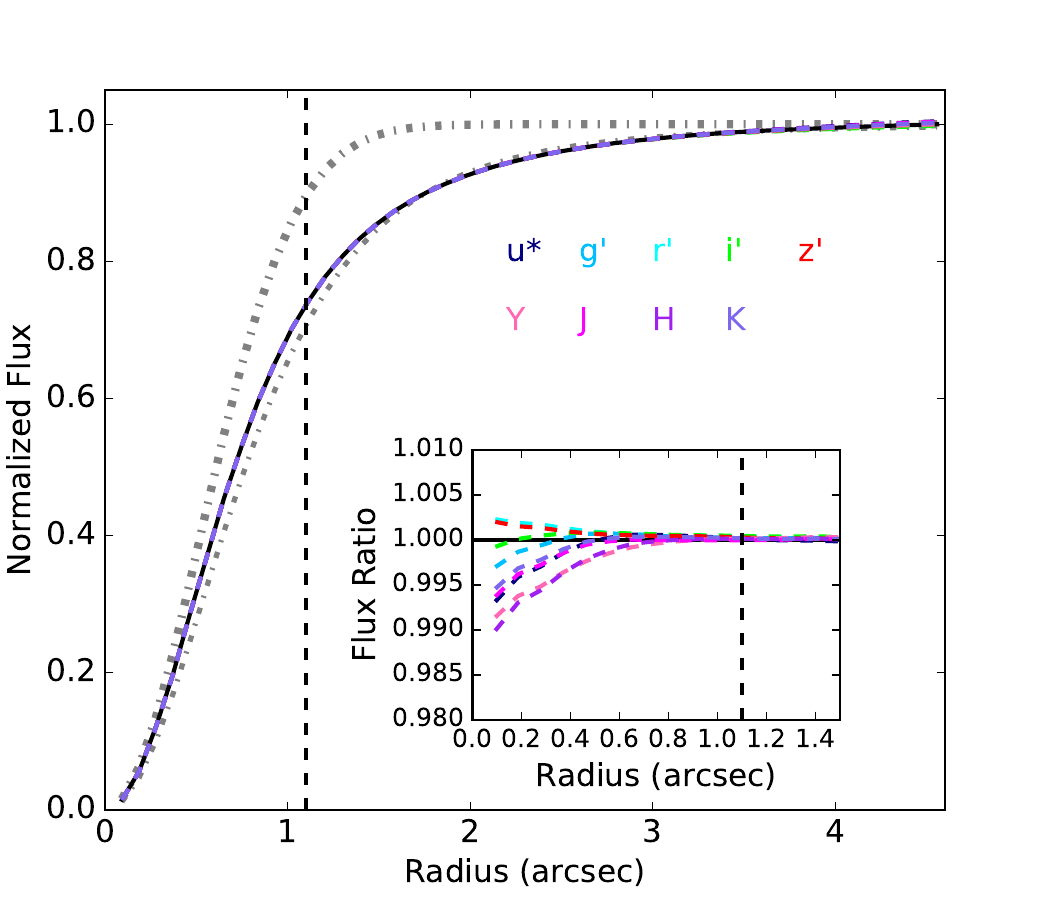}
\caption{Average curves of growth of the nine bands 
after PSF homogenization. The black solid line shows the PSF
with the worst seeing, while all the bands are displayed as color-coded 
dashed lines. The vertical dashed line is the 2.2$\arcsec$ photometric 
aperture for color measurements.
The thick and thin dashed grey lines represent the Gaussian and Moffat functions
with the same FWHM, respectively. The inserted figure shows the same curves of growth, 
but normalized to the worst one.}
\label{fig:gc}
\end{figure}

We use the \texttt{IRAF} task \texttt{lucy}, which is based on the Lucy-Richardson algorithm,
to calculate the convolution kernels to convolve with the images. All images 
are then degraded to the PSF size of $1.33\arcsec$. Figure \ref{fig:gc}
shows the average curves of growth of the PSFs after homogenization for the nine band images. 
For comparison, the thick and thin dashed grey lines illustrate the curves of growth of 
the Gaussian and Moffat profiles with the same FWHM. Obviously, the empirical PSF is 
similar to Moffat profile, but more concentrated within radius of $1.8\arcsec$. 
We display the 2.2$\arcsec$ photometric aperture used for color and photometric redshift measurements as thick black line. 
For clarity, the inserted plot shows the corresponding ratios relative to the curve of growth with 
the largest PSF size. We can see that the residuals, resulting from the numerical approximation during 
the convolution, are less than 1.0\% even at the very central region and close to zero at large radius.
However, without the homogenization the maximum residual  
at the central region can reach to 25\%.

\begin{figure*}
\includegraphics[width=1.0\textwidth]{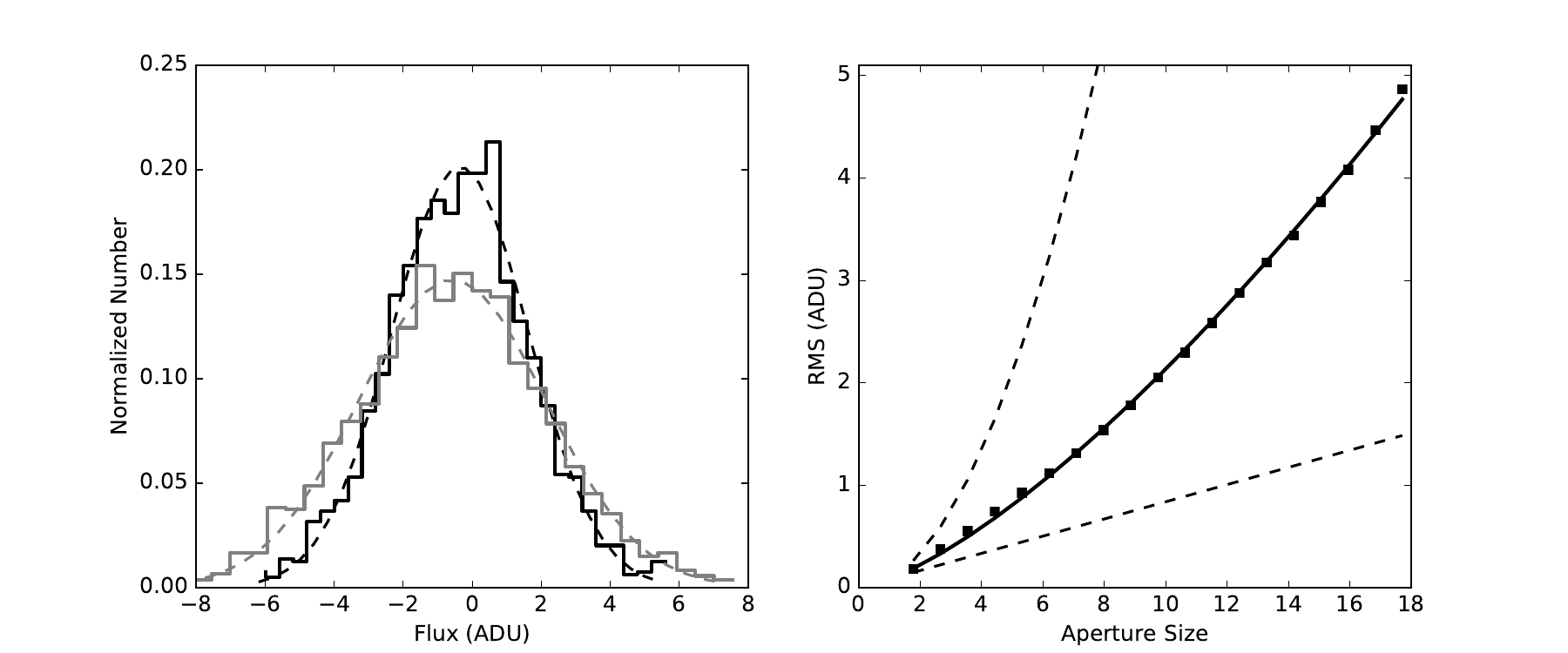}
\caption{\emph{Left panel}: histogram of number counts with aperture 
diameters of 2.0 and 2.5 arcsec. Their distributions have been normalized 
to unity. Larger aperture size shows more count variations and has 
larger Gaussian FWHM. \emph{Right panel:} Gaussian rms curve in terms 
of aperture size defined as $\sqrt{N}$. Filled squares are measured 
directly from sky-subtracted image. A power-law function is used to fit
the measurements. Meanwhile, the bottom and top dashed curves illustrate 
two limiting cases: no pixel correlation and complete correlation 
in adjacent pixels.}
\label{fig:beta}
\end{figure*}

\subsection{Source Detection and Photometry}
For each band, the PSF matched images are stacked to one mosaic,
which is about $2.1 \times 1.1$\,deg$^2$ in size. 
We then run \texttt{SExtractor} in dual-image mode for source detection and photometry. 
Due to the comparable depth to $Y$-band and good image quality, the 
non-PSF-homogenized $J$-band image is used as the detection image.
The detection threshold is set to 1.5$\sigma$ above the background and 
at least three connected pixels are required for a measurement. In total, about 120,000
objects are detected in the multi-band images.

For each object we calculate two types of magnitudes, a specific aperture magnitude 
and the Kron-aperture magnitude (\texttt{MAG\_AUTO}). In order to achieve an 
optimal signal-to-noise ratio estimate, generally, the aperture diameter is about 
1.35$\times$FWHM if the point source has Gaussian PSF. However, from Figure 
\ref{fig:gc} we know that the deviation between the empirical and Gaussian PSFs is 
very large. Therefore, the aperture diameter we use in this work for color and hence photometric redshift
measurements is 2.2$\arcsec$ which is about 1.65 times of the FWHM. To remedy the 
missing fluxes due to the limited aperture, the aperture
correction is applied based on the derived growth of curve. 
The Kron-aperture magnitude is measured within 2.5 times Kron 
radius \citep{1980ApJS...43..305K} which accounts for over 96 per cent of 
the total flux of a galaxy, and provides an accurate estimate of the total magnitude.

We correct the Galactic extinction for the measured magnitudes using the dust maps
from \citet{1998ApJ...500..525S} assuming R$_V$=3.1. The Galactic extinction curve 
is from \citet{1989ApJ...345..245C}. The final corrections are less than 4.0\% 
for optical bands and 1.0\% for the near-infrared.

\subsection{Noise Correction}
There are two kinds of sources that contribute to photometric 
error budgets: photon shot noise from observed objects, and sky background 
fluctuations. The sky background noise can be, generally, determined 
by $\sigma^2=\sigma_{0}^{2}N$, where $\sigma_{0}$ is the standard deviation 
of background noise and $N$ is the pixel area for a given aperture. However,
due to the existence of noise correlation stemming from resampling and 
PSF homogenization procedures, the standard formula will underestimate the photometric 
errors. Taking the noise correlation into account, the background noise estimate
can be generalized as $\sigma^2\propto{N^{2\beta}}$ where $\beta$ is a free 
parameter within [0.5, 1.0]. In the case of pure background noise, $\beta=0.5$, 
which returns to the conventional formula. If the adjacent pixels are completely 
correlated, on the other hand, $\beta=1.0$.

We follow the recipe presented by \citet{2007AJ....134.1103Q} to determine $\beta$ 
for each image. Similar method can also be found in \citet{2006ApJS..162....1G} and 
\citet{2003AJ....125.1107L}. We randomly generate a set of $\sim$\,1000 positions on each 
band image which do not overlap with detected objects. The fluxes 
can be measured for each position using different apertures. For a given aperture, we fit 
the distribution of the measured fluxes with a Gaussian function. Larger aperture generally has larger 
Gaussian dispersion. Then we use the power-law equation described above to fit the relation between
Gaussian dispersion and aperture size. While $\beta$ changes between different bands, 
the typical value is 0.6-0.8. Figure \ref{fig:beta} shows an example of the fitting procedure for 
$Y$-band image. The correlated noise is transferred to the final magnitude errors by following the 
equation of error adopted by \texttt{SExtractor}.

\section{Photometric Redshifts of Galaxies}\label{sec:photoz}
We measure the photometric redshifts of the detected galaxies using the Bayesian 
photometric redshift code \texttt{BPZ} \citep{2000ApJ...536..571B, 2006AJ....132..926C}.
We also apply the \texttt{EAZY}  code \citep{2008ApJ...686.1503B} for comparison
purposes. \texttt{BPZ} is a spectral 
template-based code in combination with a redshift prior derived from the 
spectroscopic redshifts in the Hubble Deep Field-North (HDF-N). 
\texttt{EAZY} is another template-fitting program which can linearly combine different templates 
and estimate realistic redshift uncertainties by introducing a novel rest-frame template error function.
These two codes are constantly updated, and have been widely used in many studies.
It should be noted that we are not aiming to compare the performance of the two codes themselves, 
but to study the characteristics of photometric redshifts derived by them.
As in \citet{2012MNRAS.421.2355H}, the recalibrated template set 
of  \citet{2004PhDT.........4C} is used in this work\footnote{As discussed in 
\citet{2008ApJ...686.1503B}, \texttt{EAZY} provides five default templates which are optimal 
for many features of the code. However, for our study we use the idential set of templates as 
\texttt{BPZ}, though it could degrade the performance of \texttt{EAZY}.}.
Before running the codes, we carry out two improvements as follows.

As discussed in \citet{2012MNRAS.421.2355H}, since the peak of the posterior probability distribution,
the product of the redshift likelihood and the Bayesian prior, is always at $z > 0$, the photometric redshift values 
estimated by \texttt{BPZ} at low redshifts are systematically overestimated. \citet{2012MNRAS.421.2355H} modified 
the prior so that it no longer vanishes for z = 0. As a result, it leads to the improved photometric redshift
estimate at low redshift. On the other hand, \citet{2014ApJ...797..102R} reconstructed the prior using
SDSS spectroscopic Galaxy Main Sample \citep{2002AJ....124.1810S, 2014ApJS..211...17A} for 
galaxies with $12.5\,\mathrm{mag} \leq i' \leq 17.0\,\mathrm{mag}$ 
and VVDS spectroscopic sample \citep{2013A&A...559A..14L} for galaxies with $i' \geq 20.0\,\mathrm{mag}$. 
For intermediate magnitudes, they interpolated the parameters of 
the prior to match those fitted at $i' = 17.0\,\mathrm{mag}$ and $i' = 20.0\,\mathrm{mag}$. 
The new prior can significantly reduce the photometric redshift bias and outliers for galaxies brighter than $20.0\,\mathrm{mag}$. 
Therefore, it improves the photometric redshift accuracy at low redshift ($z < 0.4$). 
Here we also apply this new prior derived by 
\citet{2014ApJ...797..102R} to \texttt{BPZ} for photometric redshift measurements.
As for \texttt{EAZY}, we use its default prior which is constructed with the synthetic photometry 
of galaxies in the semianalytic model \citep{2007MNRAS.375....2D}. Unlike  \texttt{BPZ},  
one major feature of \texttt{EAZY} prior is that it does not impose any color restrictions as a 
function of redshift.

Any small color offsets of a few percent between different bands can affect the photometric redshift accuracy. 
For template-fitting methods, the commonly used strategy 
is to add some zeropoint offsets with an iterative process by comparing the measured and predicted
colors (e.g. \citet{2006A&A...457..841I, 2010ApJS..189..270C, 2012MNRAS.421.2355H}). 
By analyzing CFHTLenS data, \citet{2012MNRAS.421.2355H} concluded that such correction can
suppress the PSF effects. Nevertheless, a proper PSF homogenization 
can be equivalent to the offset calibration if an accurate absolute photometric calibration is also performed.
Our catalog, however, has two major differences compared to CFHTLenS data. First, it includes 
photometry from a number of surveys which span a period of many years and have different data 
reduction and calibration procedures. Secondly, we assume a constant PSF across a certain field 
without considering PSF variations. Therefore, it is essential to apply the offset calibration.
We calculate the zeropoint offsets for different bands with the spectroscopic 
redshift sample described in Section \ref{sec:phzcom} by fixing the redshifts to the spectroscopic values.
For both codes, the derived absolute offset is about 0.11\,mag for $J$ band, and less than 
0.06\,mag for any other band.

\begin{figure}
\includegraphics[width=0.46\textwidth]{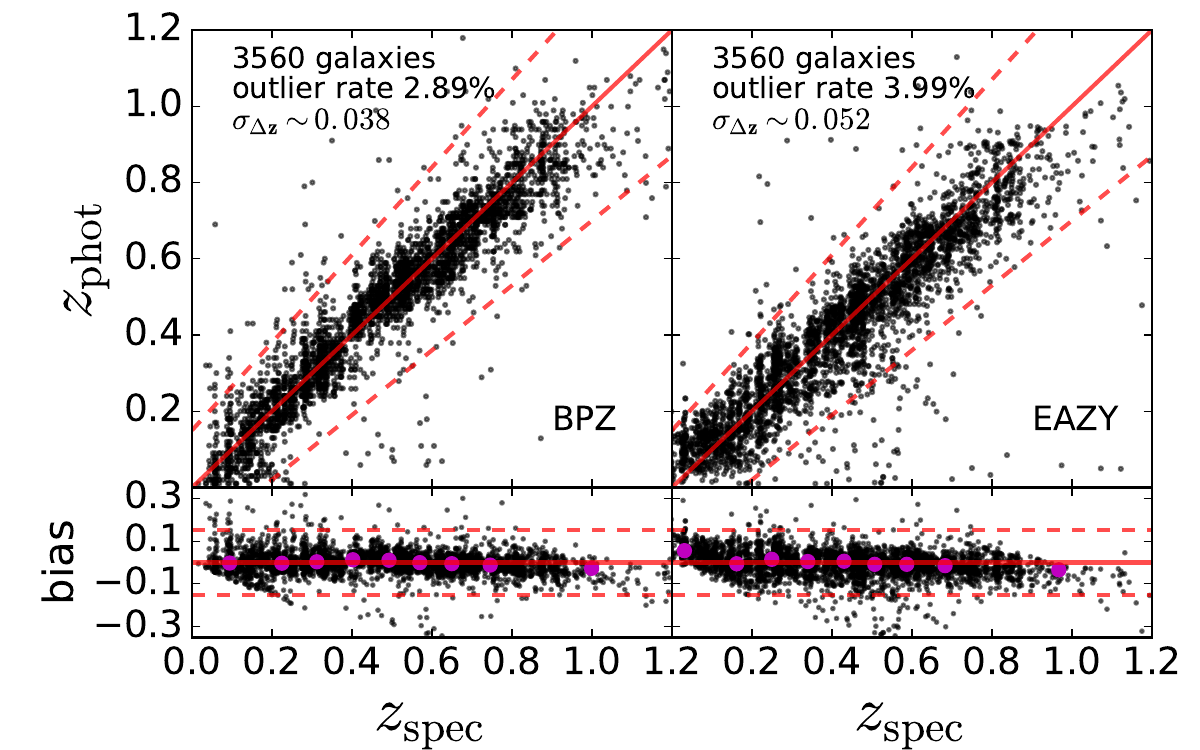}
\caption{\textit{Top panel:} comparison between the spectroscopic and photometric redshifts 
with $i'$-band magnitude at $16.5\,\mathrm{mag}<i'<24.0\,\mathrm{mag}$. 
The red dashed lines represent the outlier limit
defined as $z_\mathrm{p} = z_\mathrm{s} \pm 0.15\times(1.0+z_\mathrm{s})$. 
\textit{Bottom panel:} bias distribution as a function of spectroscopic redshifts. The magenta dots 
are the median values in eight redshift bins.}
\label{fig:photoz}
\end{figure}

\begin{figure}
\includegraphics[width=0.45\textwidth]{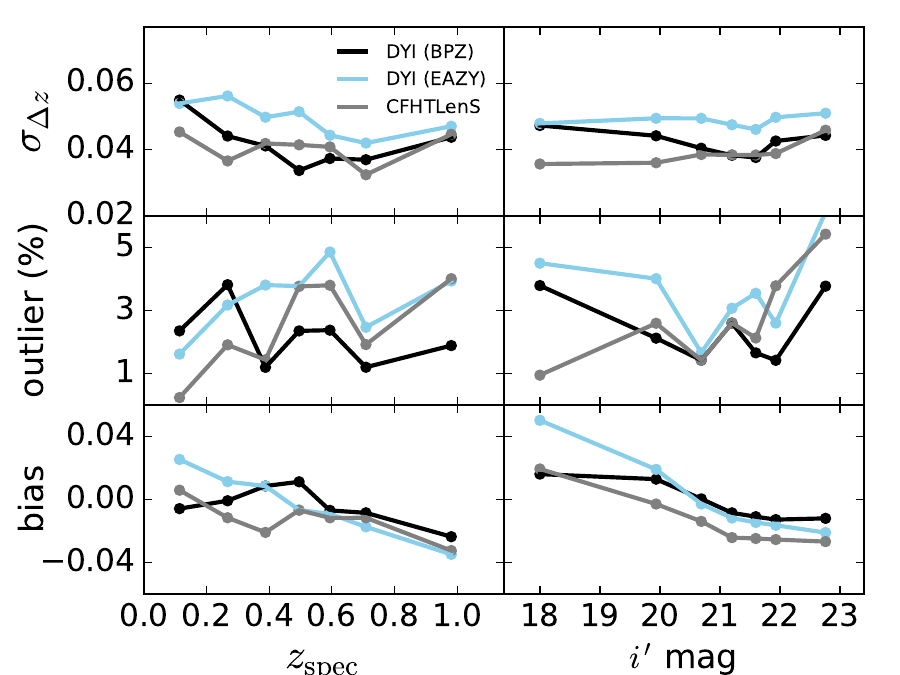}
\caption{Photometric redshift statistics as a function of redshift (\textit{left}) and $i'$-band magnitude (\textit{right}).
Our results are shown as black (\texttt{BPZ}) and light blue (\texttt{EAZY}) lines, while CFHTLenS results 
are in grey. The magnitudes and redshifts
are binned so that there are equal numbers of galaxies in each bin.}
\label{fig:photoz_cfht}
\end{figure}

\subsection{Photometric Redshift Accuracy}\label{sec:phzcom}
To check the accuracy of the photometric redshift measurements, we collect the spectroscopic redshifts of galaxies
from three catalogs: the final data release of VVDS  \citep{2013A&A...559A..14L},
the first data release of VIPERS \citep{2014A&A...562A..23G},
and the SDSS DR12 spectra catalog \citep{2015ApJS..219...12A}.
Among the catalogs, only galaxies with the most secure redshift measurements are selected. 
Specifically, galaxies with flag 3 and 4 (confidence level above 95\%) are used for 
both VVDS and VIPERS data, while the SDSS galaxies are included to have spectra with \texttt{zWarning} 
equal to zero. For the duplicated galaxies the selection priority is VIPERS, VVDS and SDSS. 
In total, we obtain a sample of 3560 secure spectroscopic redshifts over the entire field. Their $i'$-band magnitudes
range from 16.5 to 24.0\,mag, while the redshifts are in the range $0.02<z<1.5$ with median of $z=0.5$.
Their positions are shown in Figure \ref{fig:dyian} as blue crosses.

We access the photometric redshift accuracy by comparing to the full spectroscopic sample. 
Following other studies (e.g. \citet{2012MNRAS.421.2355H}), we define the bias as 
$\Delta{z}=\delta{z}/(1.0+z_\mathrm{spec})$, where $\delta{z}=z_\mathrm{phot}-z_\mathrm{spec}$.
Galaxies with $|\Delta{z}|>0.15$ are deemed as outliers. The Gaussian standard deviation of the bias 
is calculated after rejecting the outliers. Figure \ref{fig:photoz} shows the comparison between spectroscopic 
and photometric redshifts from the two codes. Overall, they display an satisfactory correspondence over the entire redshift range.
The percentage of the outliers is about $2.89\%$ and $3.99\%$ for \texttt{BPZ} and \texttt{EAZY}, respectively. 
The $\Delta{z}$ distribution without including outliers can be well fitted by a Gaussian function, and the derived dispersion 
is $\sigma_{\Delta{z}} \sim 0.038$ and $0.052$ for the two codes, respectively. For the photometric redshifts from 
\texttt{EAZY}, the scatters and the outlier rates are somewhat larger than those of \texttt{BPZ}. But the two results 
do not show systematic biases.
We further perform the same statistics for CFHTLenS photometric redshifts, 
derived with \texttt{BPZ}, using the spectroscopic sample.
Due to the existence of the large mask regions around saturated stars in CFHTLenS catalog, there are a total of 
2962 cross-matched galaxies. As a result, the outlier rate and standard deviation are $2.40\%$ and $0.038$ 
for our  photometric redshifts estimated with \texttt{BPZ} ($3.65\%$ and $0.051$ for \texttt{EAZY}), while these are $2.70\%$ 
and $0.039$ for CFHTLenS. Though the sigma values are almost identical, our photometric redshift measurements with 
\texttt{BPZ} exhibit a somewhat lower outlier rate. 

We further analyze the photometric redshift accuracy as a function of spectroscopic redshift and $i'$-band magnitude. 
Figure \ref{fig:photoz_cfht} shows how the standard deviation, outlier rate and bias depend on the 
two measured quantities. The redshifts and magnitudes are binned so that there are equal numbers 
of galaxies in each bin. For comparison, the statistics for CFHTLenS are illustrated as grey lines.
For redshift below 0.4 and magnitude brighter than 20.0\,mag, our photometric redshift measurements with \texttt{BPZ}
exhibit relatively large $\sigma_{\Delta{z}}$ and outlier rate. However, the outlier rate shows better performance
for galaxies at $z>0.4$ and faint magnitudes ($i'\gtrsim21.0$\,mag). Furthermore, the bias is smaller than 
that of CFHTLenS almost over the entire redshift and magnitude ranges. Table \ref{tab:zcom} provides 
more detailed comparison in different spectroscopic redshift bins. The median value of the 95\% confidence intervals 
in a given bin is calculated to characterize the statistical error of the Bayesian posterior probability. 
The \texttt{BPZ} \texttt{ODDS} parameter is another quantity to describe the unimodality of a galaxy's 
posterior redshift distribution, and we 
show the fraction of galaxies with \texttt{ODDS}$\ge$0.9 in Table \ref{tab:zcom}. Significant improvements
of the photometric redshift accuracy can be seen by comparing the two catalogs.
In summary, in the redshift range $0.0<z_\mathrm{spec}<1.5$ at $i'\lesssim24.0$\,mag, the catastrophic 
fraction of our photometric redshifts estimated with \texttt{BPZ} is no more than 4.0\%, and the scatter is in the range 
$0.03<\sigma_{\Delta{z}}<0.06$.

\begin{figure*}
\centering
\includegraphics[width=0.95\textwidth]{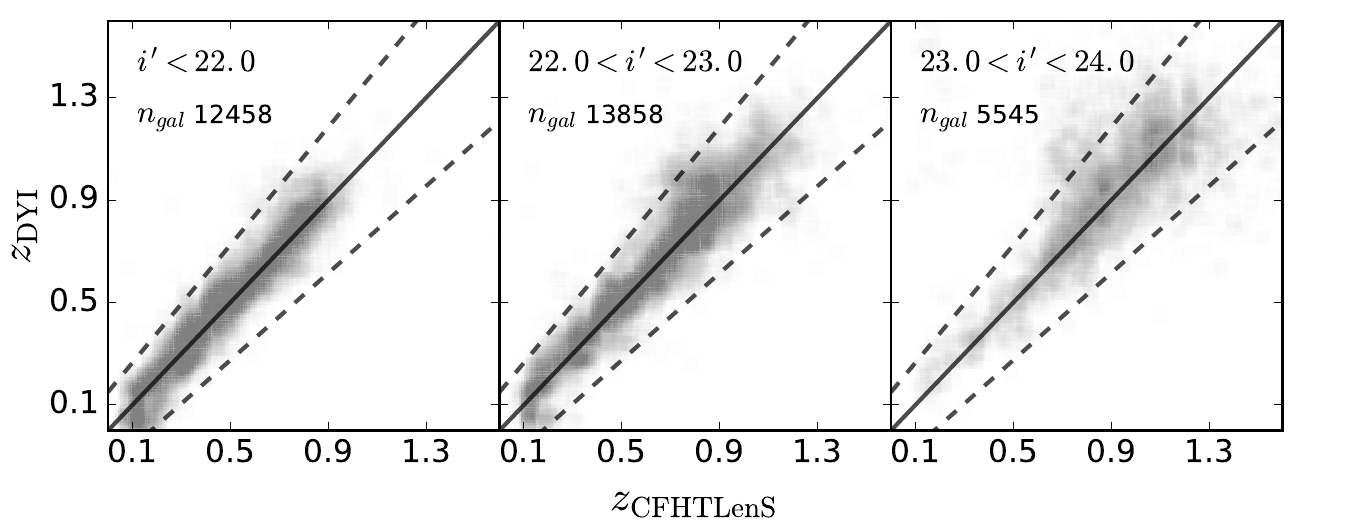}
\caption{Comparison of the photometric redshifts measured by both our DYI (\texttt{BPZ}) and CFHTLenS
in three $i'$-band magnitude bins. The galaxies are selected to have \texttt{ODDS}$>$0.9 
and \texttt{SExtractor FLAGS=0} in both samples ($\sim$\,30,000 galaxies in total). 
The black dashed lines represent the outlier limit.}
\label{fig:photoz_cfht2}
\end{figure*}

Figure \ref{fig:photoz_cfht2} compares our photometric redshift results of galaxies with those of CFHTLenS. 
To ensure the comparison to be conclusive, we only use the photometric redshifts estimated 
with \texttt{BPZ} and restrict the galaxies to have \texttt{ODDS} values larger than 0.9
and \texttt{SExtractor FLAGS=0} in both samples. With these criteria, most of the galaxies at redshift 
$z\gtrsim1.3$ in CFHTLenS sample are rejected because of the large photometric redshift uncertainties.
We split the $i'$-band magnitudes into three bins: $i'<22.0$\,mag, $22.0\,\mathrm{mag}<i'<23.0\,\mathrm{mag}$,  
and $23.0\,\mathrm{mag}<i'<24.0\,\mathrm{mag}$. The number of galaxies in each bin is listed in the corresponding
panels. The figure shows an sharp decrease for $z_\mathrm{CFHTLenS}$ 
at $z\sim0.1$ for $i'<23.0$\,mag. As stated in \citep{2012MNRAS.421.2355H}, 
it is mainly attributed to the selected redshift prior. For $z\gtrsim0.2$, our photometric redshift measurements show
unbiased correspondence with $z_\mathrm{CFHTLenS}$ (scatter $\sim$\,0.039, outlier rate $\sim$\,3.82\%). 
However, for fainter galaxies ($23.0\,\mathrm{mag}<i'<24.0\,\mathrm{mag}$) at redshift $z>0.6$, 
the photometric redshift scatter and outlier rate 
quickly increase to 0.06 and 11.4\%. Taking the results presented in Figure \ref{fig:photoz_cfht} into account, 
our photometric redshifts are expected to be more accurate at high 
redshift due to the inclusion of near-infrared bands.

\begin{deluxetable}{lr | rcrrr}
\tablewidth{0pt}
\tablecaption{Photometric Redshift (Estimated with \texttt{BPZ}) Quality Versus Redshift. \label{tab:zcom}}
\tablehead{
\colhead{$z_\mathrm{spec}$} &  \multicolumn{1}{c|}{$n_z$} &
\colhead{$w_{95}$\tablenotemark{a}} & \colhead{\texttt{odds}$\ge$0.9} & \colhead{bias} & \colhead{outlier} & \colhead{scatter} 
}
\startdata
\multicolumn{7}{l}{DYI} \\
\hline
$(0.0, +)  $  & 2962 &0.191& 98.5\% & -0.002 & 2.4\%& 0.038 \\
$(0.0, 0.2]$ & 378   &0.192& 96.8\% & -0.007 & 1.9\%& 0.055  \\
$(0.2, 0.4]$ & 697  &0.196& 98.6\% &  0.000 & 3.0\%& 0.043 \\
$(0.4, 0.6]$ & 840  &0.192& 99.5\% &  0.010 & 2.1\%& 0.037 \\
$(0.6, 0.8]$ & 679  &0.189& 98.2\% & -0.011 & 1.3\%& 0.037 \\
$(0.8, 1.0]$ & 312  &0.187& 99.0\% & -0.021 & 1.3\%& 0.043 \\
$(1.0, +)  $  &  56   &0.177& 94.6\% & -0.079 &19.6\%& 0.042  \\
\hline
\multicolumn{7}{l}{Without $Y$ band} \\
\hline
$(0.0, +)  $  & 2962 &0.190& 98.6\% & -0.007 & 2.4\% & 0.039  \\
$(0.0, 0.2]$ & 378   &0.190& 98.9\% & -0.016 & 1.1\% & 0.052  \\
$(0.2, 0.4]$ & 697  &0.194& 98.6\% & -0.008 & 3.2\% & 0.043  \\
$(0.4, 0.6]$ & 840  &0.191& 98.9\% &  0.007 & 2.3\% & 0.038 \\
$(0.6, 0.8]$ & 679  &0.189& 98.1\% & -0.012 & 1.8\% & 0.038 \\
$(0.8, 1.0]$ & 312  &0.186& 99.0\% & -0.024 & 1.3\% & 0.044 \\
$(1.0, +)  $  &  56   &0.176& 96.4\% & -0.085 &19.6\%& 0.044 \\
\hline
\multicolumn{7}{l}{CFHTLenS} \\
\hline
$(0.0, +)  $  & 2962 &0.271& 79.9\% & -0.013 & 2.7\% &  0.039 \\
$(0.0, 0.2]$ & 378   &0.262& 89.9\% &   0.007 & 0.3\% &  0.045 \\
$(0.2, 0.4]$ & 697  &0.293& 61.1\% & -0.015  & 1.3\% &  0.038 \\
$(0.4, 0.6]$ & 840  &0.267& 91.1\% & -0.010  & 3.5\% &  0.044\\
$(0.6, 0.8]$ & 679  &0.267& 83.8\% & -0.012  & 2.7\% &  0.034\\
$(0.8, 1.0]$ & 312  &0.291& 80.1\% & -0.034  & 1.9\% &  0.043\\
$(1.0, +)  $  &  56   &0.419& 30.4\% & -0.077  & 30.4\% & 0.059
\enddata
\tablenotetext{a}{The median 95\% confidence interval estimated by \texttt{BPZ}.}
\end{deluxetable}

\begin{figure*}
\centering
\includegraphics[width=0.95\textwidth]{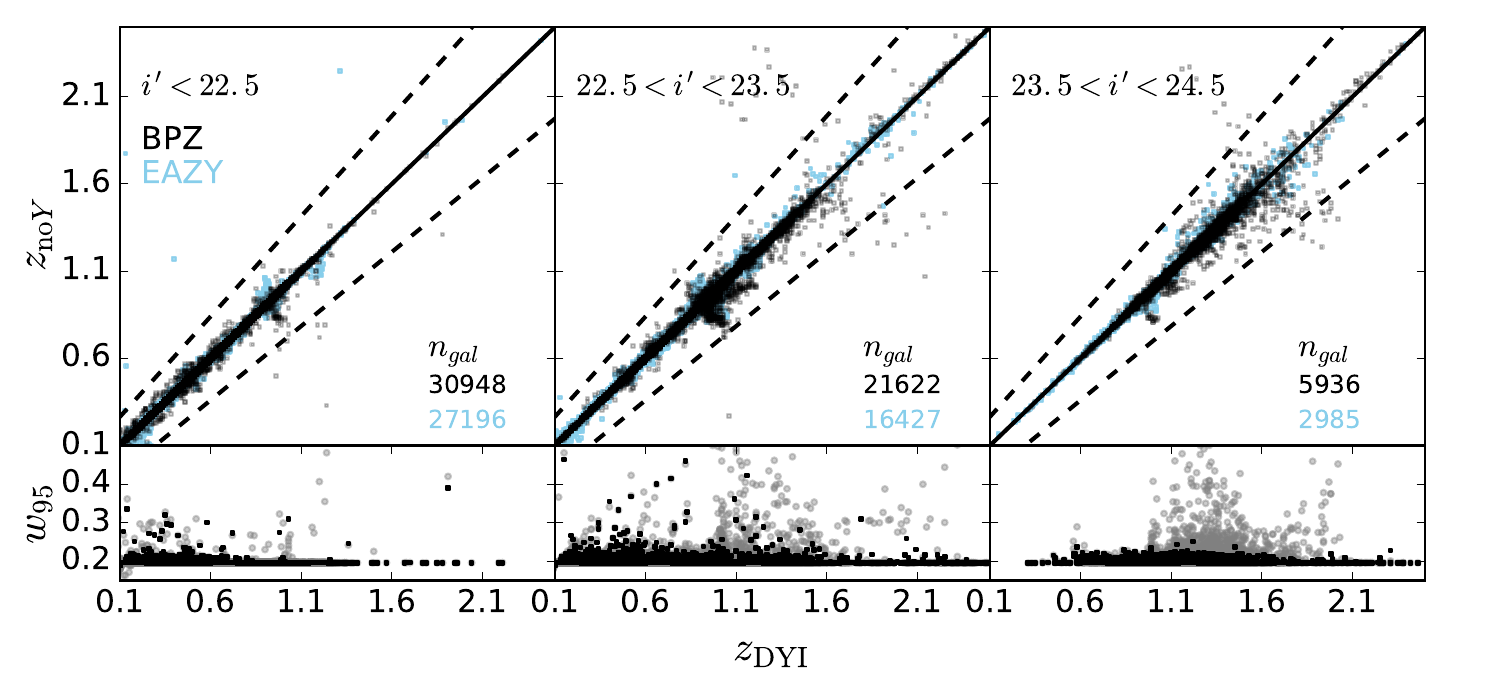}
\caption{\textit{Top panel:} comparison between the photometric redshifts without including $Y$ band 
($z_{\mathrm{no}Y}$) and the photometric redshifts estimated by all bands with hight validation in 
three $i'$-band magnitude bins. Total $\sim$\,59,000 galaxies for \texttt{BPZ} (black dots) 
and $\sim$\,47,000 for \texttt{EAZY} (light blue dots) are selected with \texttt{ODDS}$>$0.95 
and reduced $\chi^2\le1.0$. The black dashed lines represent the outlier limit.
\textit{Bottom panel:} the 95\% confidence interval as a function of redshift. Black dots indicate 
the photometric redshifts estimated with \texttt{BPZ} using all bands, while grey dots are $z_{\mathrm{no}Y}$.}
\label{fig:zp2Y}
\end{figure*}

\begin{figure*}
\centering
\includegraphics[width=0.95\textwidth]{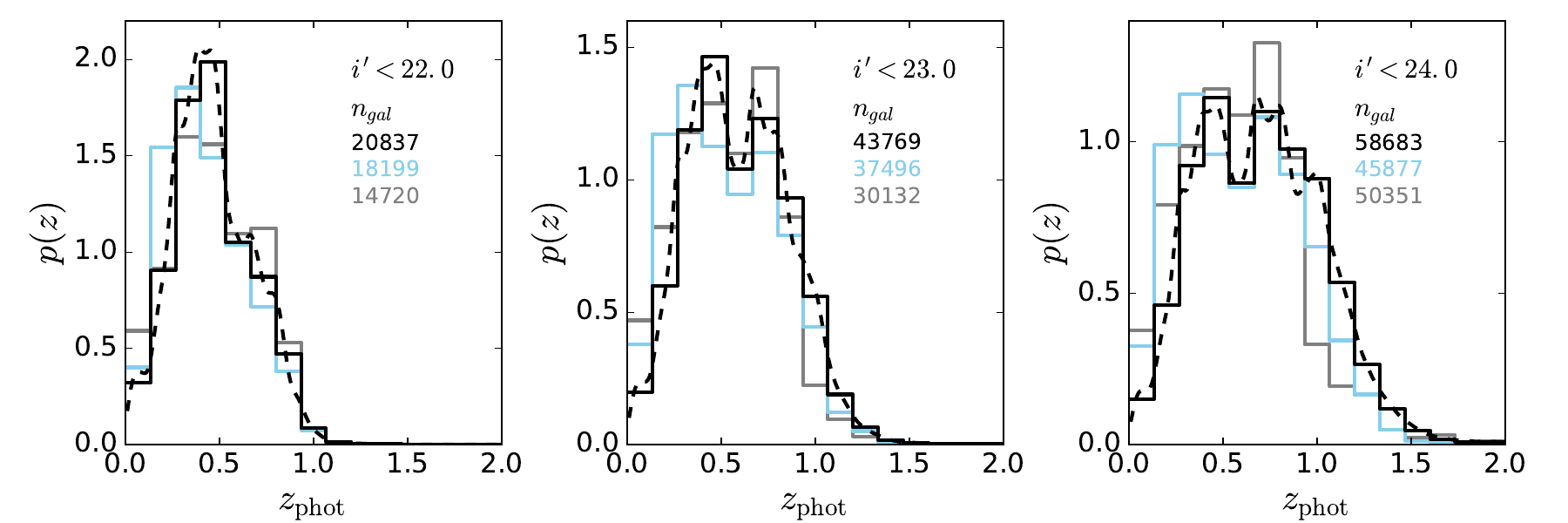}
\caption{Redshift distributions for three magnitude limits: $i'<22.0$\,mag, $i'<23.0$\,mag, and $i'<24.0$\,mag. 
The black solid line and light blue line show the distributions of DYI photometric redshifts estimated with 
\texttt{BPZ} and \texttt{EAZY}, respectively, and the black dashed line is the stacking of corresponding posterior 
probabilities calculated with \texttt{BPZ}. For comparison, grey lines display the distributions of 
CFHTLenS photometric redshifts in the same field. The galaxies in the three samples are selected with reliable 
photometric redshift measurements (i.e. \texttt{ODDS}$>$0.9 and reduced $\chi^2\le1.0$). Their numbers
in each magnitude bin are listed. Note that  there are total $\sim$\,80,000 and $\sim$\,283,000 galaxies in DYI and 
CFHTLenS samples, respectively, before applying the selection criteria.}
\label{fig:zdis}
\end{figure*}

\subsection{Photometric Redshifts Without $Y$ Band}
One of our goals is to analyze the effect of the $Y$-band photometry on galaxy's photometric redshift measurements. 
Because the dominant 4000\,{\AA}  break feature of galaxies is redshifted to $Y$ band at
redshift range $z\sim1.25-1.8$, including $Y$ band should improve the photometric redshift estimate for that range.
However, there are only very few spectroscopic redshifts larger than $z\sim1.2$ in the field (see Table \ref{tab:zcom}). 
To extend the redshift range, instead, we select a subsample of photometric redshifts estimated 
by all bands with high validation (\texttt{ODDS}$>$0.95 and reduced $\chi^2\le1.0$), and 
compare the photometric redshifts excluding the $Y$ band ($z_{\mathrm{no}Y}$) with this subsample, 
as shown in the top panel of Figure \ref{fig:zp2Y}. For \texttt{BPZ}, 
we can see that they show good correspondence with scatter less than 0.008 at $z<1.0$ in the 
three magnitude bins. For $z$ in the range $1.0<z<2.0$, where the 4000\,{\AA} break basically locates 
in $Y$-band wavelength, $z_{\mathrm{no}Y}$ presents a large scatter of about 0.017 after rejecting 
outliers. Beyond $z\sim2.0$, the scatter decreases again because the 4000\,{\AA} break has shifted to 
longer wavelengths. Similar results are also presented for photometric redshifts from \texttt{EAZY}.

In addition to degrading the photometric redshift accuracy, the exclusion of $Y$-band photometry would also reduce 
its precision. The bottom panel of Figure \ref{fig:zp2Y} presents the comparison of the 95\% 
confidence interval, denoted as $w_{95}$,  as a function of redshift. Due to the different 
definitions on the confidence level between the two photometric redshift codes, here we only show the results from 
\texttt{BPZ}, though we can draw a very similar conclusion from \texttt{EAZY}. 
The black dots in the figure indicate the photometric redshifts measured with all 
bands, while grey dots are $z_{\mathrm{no}Y}$. It is seen that the precision of $z_{\mathrm{no}Y}$ 
gets worse at $1.0<z<2.0$ as the galaxies become fainter. We arbitrarily define the fraction of 
$w_{95}>0.22$ to analyze the precision quantitively. For the three magnitude bins, 
the fraction increases from 0.2\% to 10.4\% for $z_{\mathrm{no}Y}$, while it is always less than 
0.6\% for $z_\mathrm{DYI}$. Based on the above analysis, we conclude that the inclusion of 
$Y$-band photometry can capture the 4000\,{\AA} break so that it is capable to improve the 
photometric redshift measurements at redshift range $1.0<z<2.0$.

\begin{figure*}
\centering
\includegraphics[width=0.95\textwidth]{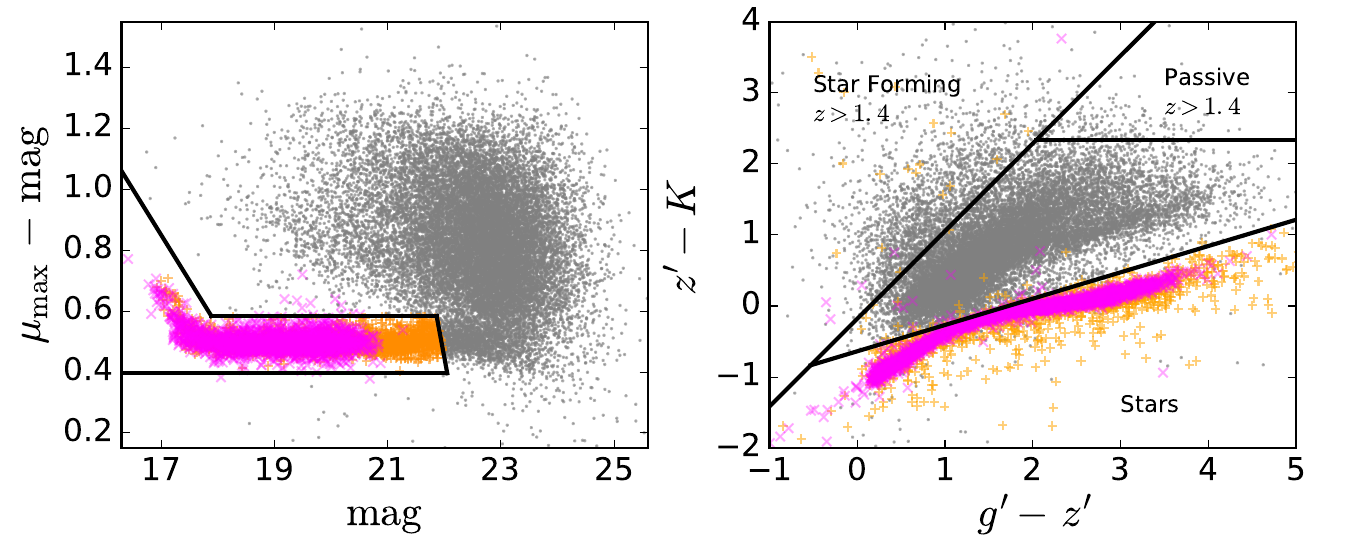}
\caption{\textit{Left panel:}  separation between the point and extended sources within the
$\mu_\mathrm{max}-\mathrm{mag}$ versus mag plane. The black line region indicates the 
locus of the point sources, shown as dark orange crosses ($+$). The two parallel lines are the $4\sigma$ limits
of the locus. The SDSS photometric standard stars are overlaid as magenta crosses ($\times$).
\textit{Right panel:} star and galaxy distributions in the $g'z'K$ color-color diagram. 
The black lines represent the classification between stars and different $BzK$ 
galaxy populations \citep{2004ApJ...617..746D}. Note that ($z'-K$) has been adjusted by -0.163\,mag
to match the relation derived by \citet{2012A&A...545A..23B} with WIRCam $K_s$ band.
The distribution of our final selected stars are shown as orange crosses ($+$).}
\label{fig:stargal}
\end{figure*}

\subsection{Redshift Distribution}
We present the redshift distributions of the measured photometric redshifts at three $i'$-band magnitude limits 
in Figure \ref{fig:zdis}. The black solid lines and light blue lines display the distributions of the reliable 
photometric redshifts estimated with \texttt{BPZ} and 
\texttt{EAZY}, respectively, and the black dashed 
lines are the corresponding stacked posterior probabilities calculated
with \texttt{BPZ}. They present good agreements in the given magnitude limits.
For comparison, we also show the distributions of CFHTLenS photometric redshifts in the same field. As excepted,
the galaxy fraction at $z>1.0$ is larger for our DYI photometric redshifts because of the more reliable measurements.
We note that there are double peaks for magnitude limits $i'<23.0$\,mag and $i'<24.0$\,mag in the 
three photometric redshift distributions. Similar behavior also presents in several other studies (e.g. 
\citet{2006A&A...457..841I,2009A&A...500..981C,2012MNRAS.421.2355H,2015MNRAS.454.3500K}). 
\citet{2012MNRAS.421.2355H} discussed that the multiple peaks probably result from the effects of filter set
and selected redshift prior. In addition, \citet{2016PhRvD..94d2005B} analyzed the photometric redshifts of
Dark Energy Survey Science Verification (DES SV) data using different photometric redshift codes. 
The resulted redshift distributions have different behaviors,  displaying either multiple peaks or single peak. 
Therefore, the presence of multiple peaks is also likely to be model-dependent.
Complete spectroscopic redshift sample down to $i'\sim24.0$\,mag and
detailed simulations are needed to address the question in detail.

\subsection{Star-Galaxy Separation}
We combine three criteria to separate stars from galaxies, the first using the maximum surface 
brightness ($\mu_\mathrm{max}$) versus magnitude diagram, the second using the $g'z'K$ color diagram 
(analog to the $BzK$ diagram presented by \citet{2004ApJ...617..746D}), and the third comparing 
the reduced $\chi^2$ derived by galaxy and stellar spectral templates. We note that the similar method 
has also been discussed by \citet{2016A&A...590A.102M}.

The bright point sources can be well separated in the $\mu_\mathrm{max}$ versus magnitude diagram 
due to the proportional relation between the light distribution of the a point source and its 
magnitude \citep{2007ApJS..172..219L}. We use $i'$-band photometry for the selection, and cut 
the maximum surface brightness to a given limit $\mu^{lim}_\mathrm{max}$. 
The locus of the point sources are within the black line region as shown in the left panel of 
Figure \ref{fig:stargal}. The selected point source sample by this criterion is denoted as $\mathbf{S_{1}}$.

The $BzK$ diagram was first proposed by \citet{2004ApJ...617..746D} to select moderate redshift 
($z\gtrsim$ 1.4) star-forming and passive galaxies. But it also turns out to be efficient for star and 
galaxy separation. \citet{2012A&A...545A..23B} transformed the equations of $BzK$ selection into 
$g'z'K_{s}$ color-color plane, where $g'z'$ are CFHT MegaPrime filters and $K_{s}$ CFHT WIRCam filter. 
Because the $K$ filter we are using is from UKIRT WFCam, an offset of -0.163\,mag is corrected 
for our ($z'-K$) color with the new conversion constant between the Vega and AB magnitude of 
WIRCam $K_{s}$ filter \citep{2014ApJS..210....4M}. The right panel of Figure \ref{fig:stargal} displays
the distribution of all objects in the $g'z'K$ diagram. The SDSS Stripe 82 photometric standard stars,
shown as magenta crosses ($\times$), are also overlaid for illustration.  
The stars are obviously separated from 
galaxies by the solid black line $(z'-K) < 0.37\times(g'-z')-0.637$. The stars selected in the $g'z'K$ diagram 
is denoted as $\mathbf{S_{2}}$.

Lastly, we rerun \texttt{BPZ} on the catalog with the stellar spectral library from \citet{1998PASP..110..863P}
by fixing the redshift to zero. The zeropoint offsets are also adjusted to fit the stellar templates. 
Then the objects are flagged as stars if they satisfy 
$\chi^2_\mathrm{star}<2\times\chi^2_\mathrm{gal}$, where the factor 2 has been set because of the 
different degrees of freedom resulting from the use of independent galaxy and stellar templates 
(e.g. \citet{2012MNRAS.421.2355H}). We denote the stars satisfied this criterion as $\mathbf{S_{3}}$.

The stars are identified finally by combining the three criteria as follows:
\begin{itemize}
\item $\mathbf{S_{1}}\cap(\mathbf{S_{2}}\cup\mathbf{S_{3}})$, for $\mu_\mathrm{max}<\mu^{lim}_\mathrm{max}$; 
\item $\mathbf{S_{2}}\cap\mathbf{S_{3}}$, for $\mu_\mathrm{max}>\mu^{lim}_\mathrm{max}$.
\end{itemize}
In addition, for objects without available photometry in one of the $g'z'K$ bands, we only
use the third criterion $\mathbf{S_{3}}$ for the classification. We select a subsample of 6236 
unsaturated stars from SDSS standard star catalog, and confirm 
that only 0.5\% stars are missed by our selection method.
Finally, about 23.9\% objects are classified as stars, and the fraction is similar to that of UKIDSS 
DXS catalog, which is $\sim$\,23.4\%, in the same field.

\subsection{Catalog}
We compile our final photometric catalog in FITS table format. In the catalog, we 
include the geometric and photometric parameters calculated by \texttt{SExtractor}.
Although several parameters (e.g. \texttt{FLAGS}, \texttt{CLASS\_STAR}, \texttt{FWHM\_IMAGE}, 
\texttt{FLUX\_RADUIS}, \texttt{MU\_MAX}, and \texttt{SNR\_WIN}) are different between 
different bands, only these in $i'$ band  are adopted.
Note that the aperture and total magnitudes are not corrected for Galactic extinction, while
the extinctions are given in separate columns. The photometric redshifts are estimated only by the galaxy spectral
templates without discriminating stars or galaxies so that the estimated redshifts of stars are not all to be zero. 
Therefore, the parameter ``type'' should be used for
star and galaxy separation in practice. The detailed description on the catalog can be seen in Table \ref{tab:cat}.  

\begin{deluxetable*}{lll}
\tablewidth{0pt}
\tablecaption{Table Contents of Our Multi-band Photometric Catalog. \label{tab:cat}}
\tablehead{\colhead{Column NO.} & \colhead{Column Name} & \colhead{Describtion}}
\startdata
1        & id                & Unique identification number, beginning from 1 \\
2, 3    & ra, dec        & \texttt{SExtractor} right ascension and declination (J2000; decimal degrees) \\
4        & type            &  Star and galaxy classification, star: type=1; galaxy: type=0 \\
5        & flag\tablenotemark{a}             & \texttt{SExtractor FLAGS} \\
6        & class\_star\tablenotemark{a}  & \texttt{SExtractor CLASS\_STAR} \\
7        & fwhm\tablenotemark{a}          & \texttt{SExtractor} FWHM assuming a Gaussian profile (pixels) \\
8, 9     & A\_image, B\_image & \texttt{SExtractor} profile rms along major and minor axes (pixel) \\
10        & theta\_image & \texttt{SExtractor} position angle (degree) \\
11        & Kron\_radius & \texttt{SExtractor} Kron aperture \\
12        & flux\_radius\tablenotemark{a}& \texttt{SExtractor} half-light radius (pixel) \\
13        & mu\_max\tablenotemark{a}    & \texttt{SExtractor} peak surface brightness (mag/arcsec$^2$) \\
14        & snr\tablenotemark{a}              &  \texttt{SExtractor} signal-to-noise ratio \texttt{SNR\_WIN} \\ 
15        & zb               & \texttt{BPZ} photometric redshift \\
16, 17  & zb\_min, zb\_max &  Lower and upper bound at 95\% confidence level of zb \\
18      & odds           & \texttt{BPZ ODDS} parameter \\
19      & chi2            &  \texttt{BPZ} reduced $\chi^2$ \\
20      & nfilt             & Number of filters used in \texttt{BPZ} for measuring photometric redshifts \\
21-29 & ext\_$x$     & Galactic extinction in the $x$ band (mag) \\
30-38 & $x$mag     &  Magnitude in the $x$ band measured within 2.2\arcsec aperture diameter \\
39-47 & $x$mag\_err & Magnitude error in the $x$ band corrected for correlation noise \\
48-56 & $x$magtot     &  Total magnitude in the $x$ band \\
57-65 & $x$magtot\_err & Total magnitude error in the $x$ band corrected for correlation noise
\enddata
\tablenotetext{a}{These parameters are measured in $i'$-band image.}
\end{deluxetable*}

\section{Summary}\label{sec:sum}
In this paper, we present a new deep CFHT WIRCam $Y$-band imaging covering about 
$2.0$ square degrees in VVDS-F22 field. Many efforts have been 
taken to properly handle various complications in image reduction and photometry. 
The final stack has reached 5$\sigma$ limit of 22.86 mag within a 2.2\arcsec\, aperture 
for point sources. The photometric dispersion, compared to UKIDSS LAS $Y$-band data, 
is 0.07\,mag. The final catalog includes broadband photometry from CFHTLenS optical images 
($u^*g'r'i'z'$) and UKIDSS DXS images ($JHK$) after PSF homogenization, totaling 
$\sim$\,80,000 galaxies.

We derive photometric redshifts using both \texttt{BPZ} package with updated redshift priors
and \texttt{EAZY}.
Comparing with spectroscopic redshifts, we find that the catastrophic fraction of our photometric redshifts 
is no more than 4.0\%, and the scatter is in the range $0.03<\sigma_{\Delta{z}}<0.06$. We also 
compare our photometric redshifts of galaxies with the corresponding CFHTLenS photometric redshifts 
derived by five optical bands in the same field. Although the two sets of results are based on different PSF 
homogenization strategies, they are generally consistent at $z>0.2$. For faint magnitude 
($23.0\,\mathrm{mag}<i'<24.0\,\mathrm{mag}$) at $z>0.6$, however, the outlier rate reaches to 11.4\%.  
In that case, our photometric redshifts are expected to be more reliable thanks to the inclusion of 
near-infrared bands. It is also confirmed by comparing with the spectroscopic redshifts. 

Many ongoing and upcoming wide field surveys, especially for 
weak lensing studies, include $Y$ band for accurate photometric redshift
measurements beyond $z\sim1.0$. We quantitatively analyze the impact of $Y$-band 
photometry on measuring photometric redshifts. Because 
there are very few spectroscopic redshifts at $z>1.2$ in the field, we select the most secure 
photometric redshifts measured by the full bands in the catalog as reference, and compare with the photometric 
redshifts without including $Y$-band photometry. For $1.0<z<2.0$, the $Y$-excluded photometric redshifts
present large scatter due to the inaccurate identification of the 4000\,{\AA} break. Our analyses verify that 
$Y$-band photometry can improve both the accuracy and the precision of photometric redshift measurements 
at $z\sim1.0-2.0$.

Our multi-band data, including images, photometry catalog and photometric redshifts, 
are available at the website: \url{http://astro.pku.edu.cn/astro/data/DYI.html}.

\acknowledgments
We would like to thank the referee for useful suggestions and 
the kind help from Weihao Wang. We also thank Mauro Stefanon
for providing some suggestions on PSF homogenization methods.
J.-Y. Yang and X.-B. Wu thank the support from the NSFC grants No.11373008 and 11533001, 
the Strategic Priority Research Program "The Emergence of Cosmological 
Structures" of the Chinese Academy of Sciences, grant No. XDB09000000, 
and the National Key Basic Research Program of China 2014CB845700.
D.-Z. Liu, S. Yuan and Z.-H. Fan  acknowledge the support from NSFC of China 
under the grants 11333001, 11173001 and 11033005, and from the Strategic Priority
Research Program "The Emergence of Cosmological Structure" of the Chinese Academy
of Science, grant no. XDB09000000. 
This research uses data obtained through the Telescope Access Program (TAP), 
which has been funded by the National Astronomical Observatories of China, the 
Chinese Academy of Sciences (the Strategic Priority Research Program "The Emergence 
of Cosmological Structures" Grant No. XDB09000000), and the Special Fund for Astronomy 
from the Ministry of Finance.
Our work is based on observations obtained with WIRCam, a joint project of CFHT, Taiwan, Korea, 
Canada, France, at the Canada-France-Hawaii Telescope (CFHT) which is operated by 
the National Research Council (NRC) of Canada, the Institut National des Sciences de 
l'Univers of the Centre National de la Recherche Scientifique of France, and the University 
of Hawaii. This work is based in part on data obtained as part of the UKIRT Infrared Deep Sky Survey.
This work is based on observations obtained with MegaPrime/MegaCam, a joint project of 
CFHT and CEA/IRFU, at the Canada-France-Hawaii Telescope (CFHT) which is operated 
by the National Research Council (NRC) of Canada, the Institut National des Sciences 
de l'Univers of the Centre National de la Recherche Scientifique (CNRS) of France, and the 
University of Hawaii. This research used the facilities of the Canadian Astronomy Data Centre 
operated by the National Research Council of Canada with the support of the Canadian Space 
Agency. CFHTLenS data processing was made possible thanks to significant computing support 
from the NSERC Research Tools and Instruments grant program. 
We acknowledge the use of SDSS spectroscopic data. Funding for SDSS-III has been provided
by the Alfred P. Sloan Foundation, the Participating Institutions, 
the National Science Foundation, and the U.S. Department of Energy Office of Science. The SDSS-III 
web site is \url{http://www.sdss3.org/}. SDSS-III is managed by the Astrophysical Research Consortium for 
the Participating Institutions of the SDSS-III Collaboration including the University of Arizona, the Brazilian 
Participation Group, Brookhaven National Laboratory, Carnegie Mellon University, University of Florida, 
the French Participation Group, the German Participation Group, Harvard University, the Instituto de 
Astrofisica de Canarias, the Michigan State/Notre Dame/JINA Participation Group, Johns Hopkins University, 
Lawrence Berkeley National Laboratory, Max Planck Institute for Astrophysics, Max Planck Institute for 
Extraterrestrial Physics, New Mexico State University, New York University, Ohio State University, 
Pennsylvania State University, University of Portsmouth, Princeton University, the Spanish Participation 
Group, University of Tokyo, University of Utah, Vanderbilt University, University of Virginia, University of 
Washington, and Yale University. This research uses data from the VIMOS 
VLT Deep Survey, obtained from the VVDS database operated by Cesam, Laboratoire d'Astrophysique 
de Marseille, France.This paper uses data from the VIMOS Public Extragalactic Redshift Survey (VIPERS). 
VIPERS has been performed using the ESO Very Large Telescope, under the "Large Programme" 182.A-0886. 
The participating institutions and funding agencies are listed at \url{http://vipers.inaf.it}. 

\facilities{CFHT (MegaPrime and WIRCam), UKIRT (WFCAM)}

\software{
BPZ \citep{2000ApJ...536..571B, 2006AJ....132..926C},
cosmics.py,
EAZY \citep{2008ApJ...686.1503B}, 
Python,
SCAMP (version 2.2.6, \citet{2006ASPC..351..112B}),
SExtractor (version 2.19.5; \citet{1996A&AS..117..393B}), 
SWarp (version 2.38.0; \citet{2002ASPC..281..228B})
}


\begin{thebibliography}{}

\bibitem[Ahn et al.(2014)]{2014ApJS..211...17A} Ahn, C.~P., Alexandroff, R., Allende Prieto, C., et al.\ 2014, \apjs, 211, 17

\bibitem[Alam et al.(2015)]{2015ApJS..219...12A} Alam, S., Albareti, F.~D., Allende Prieto, C., et al.\ 2015, \apjs, 219, 12


\bibitem[Annis et al.(2014)]{2014ApJ...794..120A} Annis, J., Soares-Santos, M., Strauss, M.~A., et al.\ 2014, \apj, 794, 120 

\bibitem[Ben{\'{\i}}tez(2000)]{2000ApJ...536..571B} Ben{\'{\i}}tez, N.\ 2000, \apj, 536, 571

\bibitem[Bertin \& Arnouts(1996)]{1996A&AS..117..393B} Bertin, E., \& Arnouts, S.\ 1996, \aaps, 117, 393

\bibitem[Bertin et al.(2002)]{2002ASPC..281..228B} Bertin, E., Mellier, Y., Radovich, M., et al.\ 2002, Astronomical Data Analysis Software and Systems XI, 281, 228

\bibitem[Bertin(2006)]{2006ASPC..351..112B} Bertin, E.\ 2006, Astronomical Data Analysis Software and Systems XV, 351, 112 

\bibitem[Bielby et al.(2012)]{2012A&A...545A..23B} Bielby, R., Hudelot, P., McCracken, H.~J., et al.\ 2012, \aap, 545, A23

\bibitem[Bonnett et al.(2016)]{2016PhRvD..94d2005B} Bonnett, C., Troxel, M.~A., Hartley, W., et al.\ 2016, \prd, 94, 042005 

\bibitem[Brammer et al.(2008)]{2008ApJ...686.1503B} Brammer, G.~B., van Dokkum, P.~G., \& Coppi, P.\ 2008, \apj, 686, 1503-1513

\bibitem[Buton et al.(2013)]{2013A&A...549A...8B} Buton, C., Copin, Y., Aldering, G., et al.\ 2013, \aap, 549, A8

\bibitem[Cardamone et al.(2010)]{2010ApJS..189..270C} Cardamone, C.~N., van Dokkum, P.~G., Urry, C.~M., et al.\ 2010, \apjs, 189, 270

\bibitem[Capak(2004)]{2004PhDT.........4C} Capak, P.~L.\ 2004, Ph.D.~Thesis, 3497 

\bibitem[Cardelli et al.(1989)]{1989ApJ...345..245C} Cardelli, J.~A., Clayton, G.~C., \& Mathis, J.~S.\ 1989, \apj, 345, 245 

\bibitem[Coe et al.(2006)]{2006AJ....132..926C} Coe, D., Ben{\'{\i}}tez, N., S{\'a}nchez, S.~F., et al.\ 2006, \aj, 132, 926


\bibitem[Condon et al.(1998)]{1998AJ....115.1693C} Condon, J.~J., Cotton, W.~D., Greisen, E.~W., et al.\ 1998, \aj, 115, 1693

\bibitem[Coupon et al.(2009)]{2009A&A...500..981C} Coupon, J., Ilbert, O., Kilbinger, M., et al.\ 2009, \aap, 500, 981 


\bibitem[Cross et al.(2012)]{2012A&A...548A.119C} Cross, N.~J.~G., Collins, R.~S., Mann, R.~G., et al.\ 2012, \aap, 548, A119 

\bibitem[Daddi et al.(2004)]{2004ApJ...617..746D} Daddi, E., Cimatti, A., Renzini, A., et al.\ 2004, \apj, 617, 746


\bibitem[De Lucia \& Blaizot(2007)]{2007MNRAS.375....2D} De Lucia, G., \& Blaizot, J.\ 2007, \mnras, 375, 2

\bibitem[Dye et al.(2006)]{2006MNRAS.372.1227D} Dye, S., Warren, S.~J., Hambly, N.~C., et al.\ 2006, \mnras, 372, 1227

\bibitem[Erben et al.(2013)]{2013MNRAS.433.2545E} Erben, T., Hildebrandt, H., Miller, L., et al.\ 2013, \mnras, 433, 2545

\bibitem[Garilli et al.(2008)]{2008A&A...486..683G} Garilli, B., Le F{\`e}vre, O., Guzzo, L., et al.\ 2008, \aap, 486, 683 

\bibitem[Garilli et al.(2014)]{2014A&A...562A..23G} Garilli, B., Guzzo, L., Scodeggio, M., et al.\ 2014, \aap, 562, A23


\bibitem[Hewett et al.(2006)]{2006MNRAS.367..454H} Hewett, P.~C., Warren, S.~J., Leggett, S.~K., \& Hodgkin, S.~T.\ 2006, \mnras, 367, 454

\bibitem[Heymans et al.(2012)]{2012MNRAS.427..146H} Heymans, C., Van Waerbeke, L., Miller, L., et al.\ 2012, \mnras, 427, 146 

\bibitem[High et al.(2010)]{2010PASP..122..722H} High, F.~W., Stubbs, C.~W., Stalder, B., Gilmore, D.~K., \& Tonry, J.~L.\ 2010, \pasp, 122, 722

\bibitem[Hildebrandt et al.(2012)]{2012MNRAS.421.2355H} Hildebrandt, H., Erben, T., Kuijken, K., et al.\ 2012, \mnras, 421, 2355

\bibitem[Ilbert et al.(2006)]{2006A&A...457..841I} Ilbert, O., Arnouts, S., McCracken, H.~J., et al.\ 2006, \aap, 457, 841 

\bibitem[Ilbert et al.(2009)]{2009ApJ...690.1236I} Ilbert, O., Capak, P., Salvato, M., et al.\ 2009, \apj, 690, 1236 

\bibitem[Ivezi{\'c} et al.(2007)]{2007AJ....134..973I} Ivezi{\'c}, {\v Z}., Smith, J.~A., Miknaitis, G., et al.\ 2007, \aj, 134, 973

\bibitem[Gawiser et al.(2006)]{2006ApJS..162....1G} Gawiser, E., van Dokkum, P.~G., Herrera, D., et al.\ 2006, \apjs, 162, 1 

\bibitem[Kron(1980)]{1980ApJS...43..305K} Kron, R.~G.\ 1980, \apjs, 43, 305 

\bibitem[Kuijken et al.(2015)]{2015MNRAS.454.3500K} Kuijken, K., Heymans, C., Hildebrandt, H., et al.\ 2015, \mnras, 454, 3500

\bibitem[Labb{\'e} et al.(2003)]{2003AJ....125.1107L} Labb{\'e}, I., Franx, M., Rudnick, G., et al.\ 2003, \aj, 125, 1107

\bibitem[Laigle et al.(2016)]{2016ApJS..224...24L} Laigle, C., McCracken, H.~J., Ilbert, O., et al.\ 2016, \apjs, 224, 24

\bibitem[Laureijs et al.(2011)]{2011arXiv1110.3193L} Laureijs, R., Amiaux, J., Arduini, S., et al.\ 2011, arXiv:1110.3193

\bibitem[Lawrence et al.(2007)]{2007MNRAS.379.1599L} Lawrence, A., Warren, S.~J., Almaini, O., et al.\ 2007, \mnras, 379, 1599 

\bibitem[Leauthaud et al.(2007)]{2007ApJS..172..219L} Leauthaud, A., Massey, R., Kneib, J.-P., et al.\ 2007, \apjs, 172, 219

\bibitem[Le F{\`e}vre et al.(2013)]{2013A&A...559A..14L} Le F{\`e}vre, O., Cassata, P., Cucciati, O., et al.\ 2013, \aap, 559, A14

\bibitem[LSST Science Collaboration et al.(2009)]{2009arXiv0912.0201L} LSST Science Collaboration, Abell, P.~A., Allison, J., et al.\ 2009, arXiv:0912.0201

\bibitem[Maddox et al.(2008)]{2008MNRAS.386.1605M} Maddox, N., Hewett, P.~C., Warren, S.~J., \& Croom, S.~M.\ 2008, \mnras, 386, 1605

\bibitem[Moutard et al.(2016)]{2016A&A...590A.102M} Moutard, T., Arnouts, S., Ilbert, O., et al.\ 2016, \aap, 590, A102

\bibitem[Mu{\~n}oz et al.(2014)]{2014ApJS..210....4M} Mu{\~n}oz, R.~P., Puzia, T.~H., Lan{\c c}on, A., et al.\ 2014, \apjs, 210, 4 

\bibitem[Pickles(1998)]{1998PASP..110..863P} Pickles, A.~J.\ 1998, \pasp, 110, 863

\bibitem[Puget et al.(2004)]{2004SPIE.5492..978P} Puget, P., Stadler, E., Doyon, R., et al.\ 2004, \procspie, 5492, 978

\bibitem[Quadri et al.(2007)]{2007AJ....134.1103Q} Quadri, R., Marchesini, D., van Dokkum, P., et al.\ 2007, \aj, 134, 1103 

\bibitem[Raichoor et al.(2014)]{2014ApJ...797..102R} Raichoor, A., Mei, S., Erben, T., et al.\ 2014, \apj, 797, 102 

\bibitem[Ramsay et al.(1992)]{1992MNRAS.259..751R} Ramsay, S.~K., Mountain, C.~M., \& Geballe, T.~R.\ 1992, \mnras, 259, 751

\bibitem[Richards et al.(2002)]{2002AJ....123.2945R} Richards, G.~T., Fan, X., Newberg, H.~J., et al.\ 2002, \aj, 123, 2945

\bibitem[S{\'a}nchez et al.(2014)]{2014MNRAS.445.1482S} S{\'a}nchez, C., Carrasco Kind, M., Lin, H., et al.\ 2014, \mnras, 445, 1482 

\bibitem[Schlegel et al.(1998)]{1998ApJ...500..525S} Schlegel, D.~J., Finkbeiner, D.~P., \& Davis, M.\ 1998, \apj, 500, 525 

\bibitem[Skrutskie et al.(2006)]{2006AJ....131.1163S} Skrutskie, M.~F., Cutri, R.~M., Stiening, R., et al.\ 2006, \aj, 131, 1163

\bibitem[Strauss et al.(2002)]{2002AJ....124.1810S} Strauss, M.~A., Weinberg, D.~H., Lupton, R.~H., et al.\ 2002, \aj, 124, 1810 

\bibitem[Taylor et al.(2009)]{2009ApJS..183..295T} Taylor, E.~N., Franx, 
M., van Dokkum, P.~G., et al.\ 2009, \apjs, 183, 295

\bibitem[van Dokkum(2001)]{2001PASP..113.1420V} van Dokkum, P.~G.\ 2001, \pasp, 113, 1420 

\bibitem[Warren et al.(2000)]{2000MNRAS.312..827W} Warren, S.~J., Hewett, P.~C., \& Foltz, C.~B.\ 2000, \mnras, 312, 827

\bibitem[Wu \& Jia(2010)]{2010MNRAS.406.1583W} Wu, X.-B., \& Jia, Z.\ 2010, \mnras, 406, 1583

\bibitem[York et al.(2000)]{2000AJ....120.1579Y} York, D.~G., Adelman, J., 
Anderson, J.~E., Jr., et al.\ 2000, \aj, 120, 1579

\end{thebibliography}
\end{document}